\documentclass[final,5p,times]{elsarticle}

\journal{Corrosion Science}

\usepackage{listings}
\usepackage[english]{babel}
\usepackage[version=3]{mhchem}
\usepackage[utf8x]{inputenc}
\usepackage{csquotes}
\usepackage[export]{adjustbox}
\usepackage{amsmath}
\usepackage{amssymb}
\usepackage{gensymb}
\usepackage{miller}
\usepackage{colortbl}
\usepackage{arydshln}
\usepackage[colorinlistoftodos]{todonotes} 
\usepackage{subcaption}
\usepackage{tabularx}
\usepackage{wasysym}
\usepackage{miller}
\usepackage{textcomp}
\usepackage{float}
\usepackage[percent]{overpic}
\usepackage{pgfplots}
\usepackage{pgfplotstable}
\usetikzlibrary{patterns}
\usetikzlibrary{positioning,pgfplots.groupplots,calc,angles,positioning,intersections,quotes,math,matrix,spy}
\usepackage{graphicx}
\usepackage[inline]{enumitem}   
\usepackage{multirow,makecell}

\makeatletter
\newcommand{\inlineitem}[1][]{%
\ifnum\enit@type=\tw@
    {\descriptionlabel{#1}}
  \hspace{\labelsep}%
\else
  \ifnum\enit@type=\z@
       \refstepcounter{\@listctr}\fi
    \quad\@itemlabel\hspace{\labelsep}%
\fi}

\usepackage{tikz}
\newcommand{\dittotikz}{%
	\tikz{
		\draw [line width=0.12ex] (-0.2ex,0) -- +(0,0.8ex)
		(0.2ex,0) -- +(0,0.8ex);
		\draw [line width=0.08ex] (-0.6ex,0.4ex) -- +(-1.5em,0)
		(0.6ex,0.4ex) -- +(1.5em,0);
	}%
}

\usetikzlibrary{arrows,decorations.pathmorphing}
\tikzset{
	arrow/.style={-stealth', line width=0.5pt},
	every picture/.append style={line width=1pt},
}

\pgfplotsset{
	compat=1.12,
	/pgf/declare function={
		f(\x,\a,\s,\c) = \a / \s /sqrt(2*pi) * exp( -0.5 * ( (\x - \c) / \s )^2 );
	},
	/pgf/declare function={
			g(\x,\a,\w,\c) = \a / \w /sqrt(pi/4/ln(2.0)) * exp( -4.0*ln(2.0) * (\x - \c)^2 / \w / \w );
	},
}

\begin{document}

\begin{frontmatter}

\title{Probing the correlation between phase evolution and growth kinetics in the oxide layers of tungsten using Raman spectroscopy and EBSD}




\author{George Fulton\corref{equal}}
\author{Artem Lunev\corref{equal}}
\cortext[equal]{These authors have contributed equally. Authors names are listed in alphabetic order.}
\ead{artem.lunev@ukaea.uk}


\address{United Kingdom Atomic Energy Authority~(UKAEA), Culham Science Centre, Abingdon, Oxfordshire OX14 3DB, United Kingdom}

\begin{abstract}
Tungsten, a plasma-facing material for future fusion reactors, may be exposed to air during abnormal operation or accidents. Only limited information is available on the evolution of related oxide phases. This work addresses the effect of substrate orientation on structural variations of tungsten oxides. Annealing experiments in an argon-oxygen atmosphere have been conducted at~$T=400$~{\textdegree}C under varying oxygen partial pressure and oxidation time. A combination of EBSD, Raman spectroscopy and confocal microscopy shows preferential oxidation initially on~\hkl{111} base material planes. The oxide scale changes its phase composition dynamically, influencing the kinetics of its growth.
\end{abstract}


\end{frontmatter}


\section{Introduction}

Tungsten, a refractory BCC metal with high thermal conductivity~\cite{ThermalDIffusivity1} and low sputtering yield~\cite{Erosion1,SputteringYield1,Sputtering2}, has been selected as the divertor material for the International Thermonuclear Reactor~(ITER)~\cite{Divertor1,Divertor2} and is a candidate armour material for the plasma-facing components~(PFC) of the DEMOnstration Power Plant. Unfortunately, there are two major issues preventing efficient use of pure tungsten for the PFC. Firstly, the low oxidation resistance~\cite{TungstenOxidation1,TungstenOxidation2} of ITER grade~($99.94$~wt.~\% fully sintered; forged and/or swaged, cold or/and hot-rolled and stress relieved~\cite{TungstenITER1}) tungsten involves certain risks during accidents and abnormal operation when a breach of vacuum is expected~\cite{JET_AirLeak,SteamTungstenReaction}. These risks include: the release of radioactive material~\cite{LOCA3} during a loss-of-coolant accident~(LOCA) with air or steam ingress~\cite{LOCA1,LOCA2}; and the deterioration and loss of plasma stability caused by the spread of volatile oxides. Another issue pertaining to the use of tungsten in fusion applications is its high brittle-to-ductile transition temperature~\cite{Erosion1,TungstenProgress1}, which may lead to cracking during transient thermal loads in the event of plasma disruption.

Multiple attempts have been undertaken to promote oxidation resistance of tungsten-based materials by proposing several ternary~(W--Cr--Y `smart alloys'~\cite{Litnovsky_2017,Klein2018,Litnovsky_2017_2,Calvo2017}) and quaternary~(W--Mo--Cr--Pd~\cite{Quaternary1,Quaternary2,Quaternary3}) tungsten alloys. Although some results look promising, the new material is still unlikely to withstand an extended LOCA event, which could last for up to three months. For instance, the passivating Cr$_2$O$_3$ layer in a W-Cr-Y smart alloy is breached after 467~h oxidation at $1000$~{\textdegree}C~\cite{SmartFail}. As more complex compositions are considered for these alloys, it becomes harder to predict how other properties~(e.g., thermal conductivity, fracture toughness) will evolve, and whether the original benefits of using tungsten are preserved. Another trend in tungsten alloy design places emphasis on improving mechanical properties of the base material and considers a completely separate set of alloying elements~(potassium, rhenium, lanthanum oxide~\cite{Fracture1}). These two lines of thought may potentially be conflicting.

A systematic study of tungsten oxidation is still lacking, although many researchers have proposed different approaches to tackle this phenomenon~\cite{Fundamental1,TungstenOxidation2,Preferential1,PreferentialOld,habainy2018}. For instance, analysis of tungsten exposed to short oxidation at low oxygen partial pressure has been previously reported in~\cite{Fundamental1,PreferentialOld,habainy2018}. \citet{Preferential1} have observed a correlation between the substrate orientation and the oxide thickness at late stages of oxidation; the highest oxidation rates were reported for the~\hkl{001} tungsten substrate orientation. In most metallic nuclear materials, either only a very small dependence of the oxidation behaviour on different substrate orientations was found, e.g. in zirconium alloys~\cite{ZrOxidation1}, or in some cases, no variation at all was found, as is the case for uranium~\cite{UraniumOxidation}. On the other hand, in non-metallic compounds such as uranium dioxide~\cite{UraniumDioxideDissolution} a marked orientational dependence on aqueous dissolution was observed. Another distinctive feature of tungsten oxidation is the vast variety of oxide phases~\cite{WOSystem1,WOSystem2}: stable stoichiometric oxides comprised of either W(IV), W(V), or W(VI); non-stoichiometric oxides formed by a mix of atomic species with different oxidation states and metastable oxides with intermediate compositions presented by W$_n$O$_{3n - 2}$ and W$_n$O$_{3n - 1}$ series, also known as the homologous Magneli phases. The latter have been observed recently by \citet{TungstenOxidation2} during oxidation of tungsten at~$600$--$800$~{\textdegree}C for up to 100~h. A recent study~\cite{habainy2018} has provided evidence of the chemical heterogeneity of the tungsten oxide scale. In addition, tungsten trioxide (WO$_3$), the higher oxide in this system, is usually found at the very surface of the oxide~\cite{habainy2018,TungstenOxidation2}, and comes in variety of polymorphous states, including: monoclinic, triclinic, tetragonal, hexagonal, and orthorhombic. The hexagonal~WO$_3$ attracts significant attention due to its unique semiconducting properties~\cite{Semiconductor1,Semiconductor2}.  

The goal of this study is to examine how the grain orientation of the tungsten substrate affects the development of an oxide scale under a controlled atmosphere with variable oxygen partial pressure and oxidation time at $400$~{\textdegree}C. Finally, an attempt is made to identify the oxide phases developed in these conditions.

\section{Material and methods}

\subsection{Sample preparation} 

Cylindrical~(\diameter~$3$~mm~$\times$~$0.5$~mm) samples were cut by electrical discharge machining from standard quality, hot-rolled pickled $99.97$~wt.~\% pure tungsten sheet~($1$~mm~$\times$~$100$~mm~$\times$~$100$~mm) supplied by Plansee. 

Grinding and polishing was performed using an AutoMet™ 250 Grinder-Polisher. For this purpose, samples were mounted on stainless steel holders with Crystalbond™~$509$. Wet grinding was done with~SiC paper followed by fine polishing with~$3~\mu$m and~$1~\mu$m diamond suspension. The final stage of polishing involved using Buehler MasterMet~2 Non-Crystallizing Colloidal Silica. Samples were thoroughly rinsed with water to avoid colloidal silica stains and the crystal bond$^{\rm{TM}}$~509 was removed with acetone in an ultra-sonic bath. Residue acetone and any carbon contamination were removed from the samples' surface by rinsing in isopropanol, followed by deionised water. Any remaining contamination observed with low-magnification optical microscopy was removed by scrubbing with PELCO optical lens tissue.

\begin{figure}[!ht]
	\centering
	\resizebox{0.55\columnwidth}{!}{%
		\begin{tikzpicture}[scale=3.5]
			\tikzstyle{arrow} = [thick,<->,>=stealth]
			\tikzstyle{singlearrow} = [thick, ->, >=stealth]
			\def \radius {1.5};
			\def \sep {2};
			\def \indentwidth {0.25};
			\def \secondindentdown {0.1};
			\def \secondindentsize {0.1};
			\def \rot {-10};
			
			\draw[black, dashed, fill=gray!30!white] (0, 0) circle (\radius);
			\draw[green!75!black,very thick] ({-\sep/2 - \indentwidth/2}, 0) -- (-\sep/2, \indentwidth*0.866) -- ({-\sep/2 + \indentwidth/2}, 0) -- cycle;
			\draw[green!75!black,very thick] ({+\sep/2 - \indentwidth/2}, 0) -- (+\sep/2, \indentwidth*0.866) -- ({+\sep/2 + \indentwidth/2}, 0) -- cycle;
			\draw[green!75!black,very thick] ({-\sep/2 - \secondindentsize/2}, -\secondindentdown-\secondindentsize*0.866) -- (-\sep/2, -\secondindentdown) -- ({-\sep/2 + \secondindentsize/2}, -\secondindentdown-\secondindentsize*0.866) -- cycle;
			\draw[rounded corners, green!75!black, dashed, ultra thick] (-\sep/2 *0.75, -\sep/2 *0.75) rectangle (\sep/2 *0.75, \sep/2 *0.75);
			
			\draw[red, very thick, rotate around={\rot:(0, 0)}] ({-\sep/2 - \indentwidth/2}, 0) -- (-\sep/2, \indentwidth*0.866) -- ({-\sep/2 + \indentwidth/2}, 0) -- cycle;
			\draw[red, very thick, rotate around={\rot:(0, 0)}] ({+\sep/2 - \indentwidth/2}, 0) -- (+\sep/2, \indentwidth*0.866) -- ({+\sep/2 + \indentwidth/2}, 0) -- cycle;
			\draw[red, very thick, rotate around={\rot:(0, 0)}] ({-\sep/2 - \secondindentsize/2}, -\secondindentdown-\secondindentsize*0.866) -- (-\sep/2, -\secondindentdown) -- ({-\sep/2 + \secondindentsize/2}, -\secondindentdown-\secondindentsize*0.866) -- cycle;
			\draw[rounded corners, red, ultra thick, dashed,rotate around={\rot:(0, 0)}] (-\sep/2 *0.3, -\sep/2 *0.3) rectangle (\sep/2 *0.3, \sep/2 *0.3);
			
			\draw[black, pattern=north west lines, pattern color=blue] (-\sep/2 *0.26, -\sep/2 *0.26) rectangle (\sep/2 *0.26, \sep/2 *0.26);
			
			\draw [arrow] (-\sep/2, \radius/1.5) -- (\sep/2, \radius/1.5 );
			\node at (0,\radius/1.4) {\Large{Indent separation: 500~$\mu\mathrm{m}$}};
			
			\draw [arrow] (-\sep/2 *0.75, \radius/1.9) -- (\sep/2 *0.75, \radius/1.9 );
			\node[green!75!black] at (0,\radius/1.75) {\Large{EBSD image: 350~$\mu\mathrm{m}$}};
			
			\draw [arrow, rotate around={\rot:(0,0)}] (-\sep/2 *0.3, -\sep/2 *0.35 ) -- (\sep/2 *0.3, -\sep/2 *0.35 );
			\node [rotate around={\rot:(0,0)},red] at (-0.0,-\radius/3.15) {\Large{Raman image: 75~$\mu\mathrm{m}$}};
			
			\draw [singlearrow] (\sep/2,0) arc (0:\rot:\sep/2);
			\node[align=center] at (\sep/2*1.1, -\radius/4.3) {\large{Misalignment}\\ \large{angle}}; 
			
			\node at (-\sep/2*1.2, \radius*0.25) {\large{Indents}};
			
			\node[align=center,blue] at (-\radius/5.0, \radius/2.6) {\Large{Overlaid area:} \\ \Large{50~$\mu\mathrm{m}$}};
			\draw [singlearrow] (-\sep/2*0.15, \sep/2*0.15) arc (0:-\rot*7:0.3);		
		\end{tikzpicture}
	}
	\caption{\label{fig:fiducial} A scheme showing the alignment procedure for producing overlays of the EBSD~(dashed green) and Raman~(dashed red) maps. Three indents are used as fiducial markers; the misalignment of indents on the SEM image~(green) and the optical image from the Raman stage~(red) is associated with manual positioning. The overlapping square area is shown as crossed-blue.}
\end{figure}
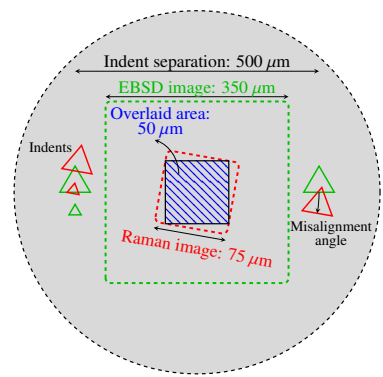

Samples were subsequently mounted on double-sided conductive adhesive copper tape. For later imaging analysis, three fiducial markers~($\approx 10$~$\mu$m in size) -- to introduce asymmetry used for orienting the sample -- were indented on each sample~(Fig.~\ref{fig:fiducial}). This was done using a Agilent Nanoindenter G200 with a Berkovich tip.

Finally, before oxidation experiments, the copper tape was removed and samples were cleaned to remove residual adhesive following the cleaning procedure above. Cleaned and oxidised samples were kept in a desiccator system at low vacuum.
     
\subsection{Characterisation methods}    

Prior to oxidation, samples were characterised by a scanning electron microscope (SEM) with electron backscatter diffraction~(EBSD) capability. In addition, large-area~($350 \ \mu\textrm{m} \times 350 \ \mu\textrm{m}$) scans were performed with confocal laser scanning microscopy (CLSM) to gather data on surface roughness. Raman spectroscopy and CLSM imaging were used to examine the oxidised layer.
 
\subsubsection{SEM and EBSD}
\label{sect:Methods_SEM_EBSD}    
    
\begin{figure}[!ht]
	\centering
	\captionsetup{singlelinecheck=off}
	\captionsetup[subfigure]{justification=centering}
	\begin{subfigure}[t]{0.5\columnwidth}
		\centering
		\subcaption{}
		\includegraphics[width=1.0\columnwidth]{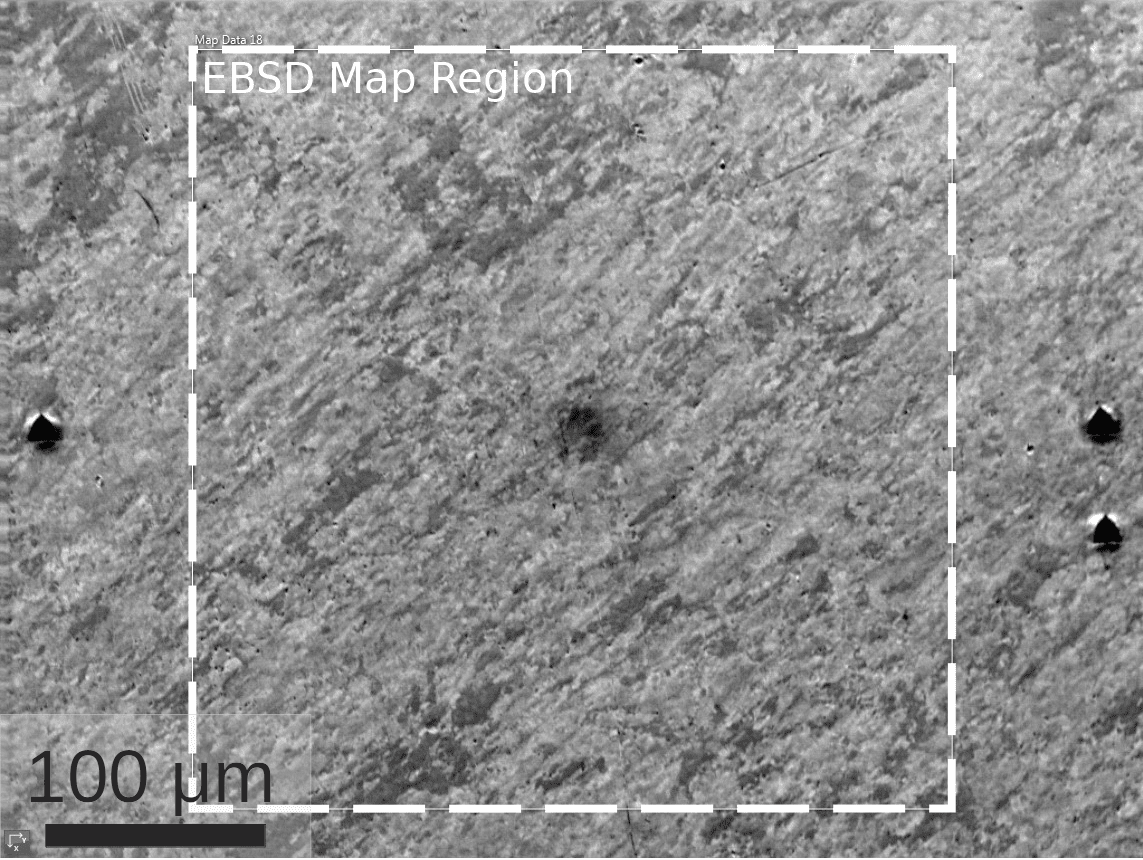}
		\label{fig:sem_example}
	\end{subfigure}%
	\qquad
	\begin{subfigure}[t]{0.3725\columnwidth}
		\centering
		\subcaption{}
		\includegraphics[width=1.0\textwidth]{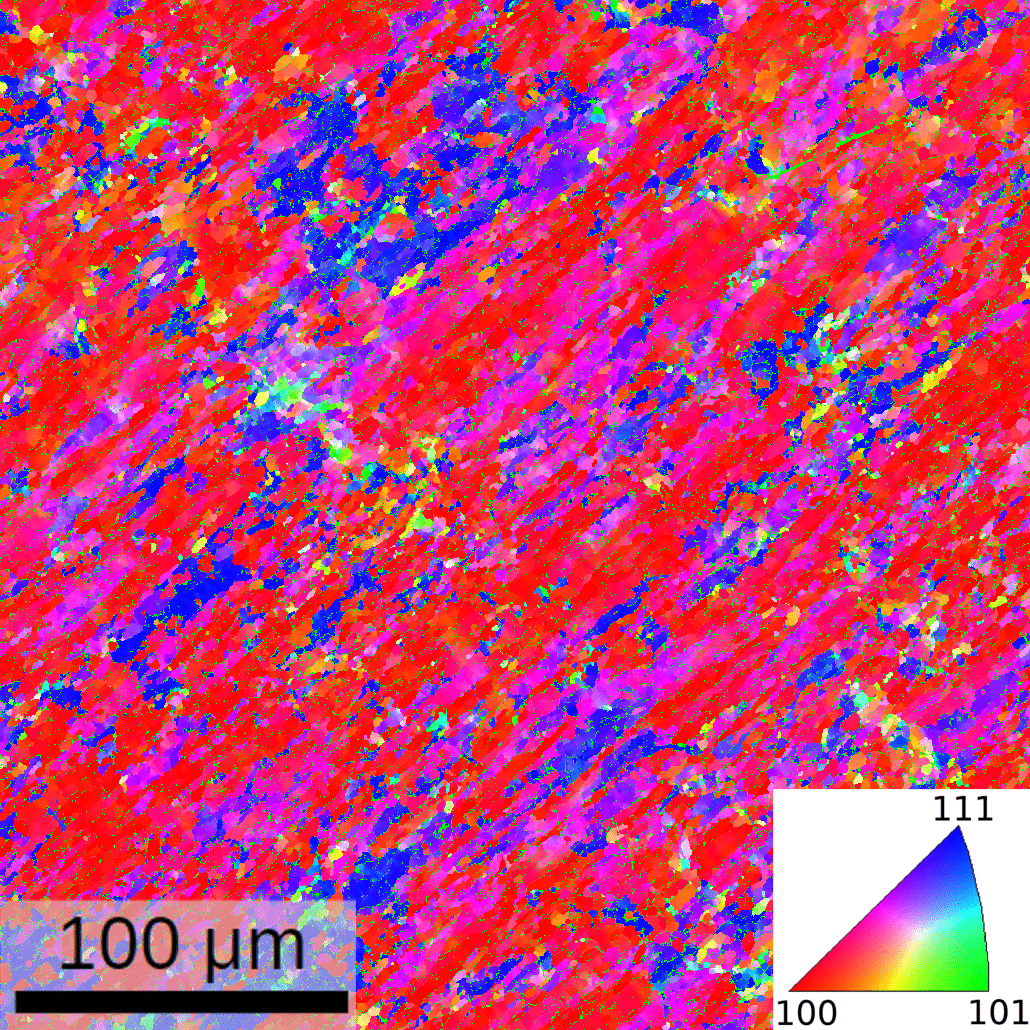}
		\label{fig:ebsd_example}
	\end{subfigure}%
	\\
	\begin{subfigure}[t]{1.0\columnwidth}
		\centering
		\subcaption{}
		\begin{tikzpicture}
		\node[inner sep = 0pt] (a) {\includegraphics[width=0.2\textwidth]{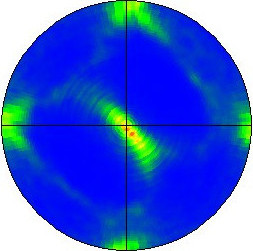}};
		\node[above=0.2cm of a,outer sep=3pt] at (a.north) {\large \hkl{001}};		
		\node[right=0.2cm of a] at (a.east) (b) {\includegraphics[width=0.2\textwidth]{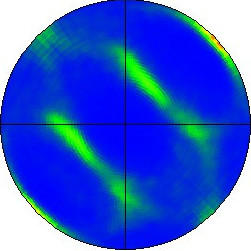}};				
		\node[above=0.05cm of b,outer sep=3pt] at (b.north) {\large \hkl{110}};
		\node[right=0.05cm of b] at (b.east) (c)		
		 {\includegraphics[width=0.2\textwidth]{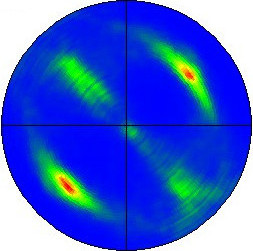}};
		\node[above=0.05cm of c,outer sep=3pt] at (c.north) {\large \hkl{111}};
		\node[right=0.2cm of c, outer sep=3pt] at (c.east) (legend) {\includegraphics[width=0.05\textwidth]{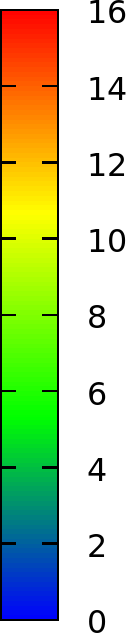}};
		\node[right=0.2cm of legend,rotate=-90] at (legend.north east) {Pole density};
		\end{tikzpicture}
		\label{fig:pole_figures_example}
	\end{subfigure}
	\caption[foo_bar]{Crystallographic texture of as-prepared samples: \begin{enumerate*}[label=(\alph*)]
			\item Mixed secondary electrons~(SE) and forward scatter detector~(FSD) image showing topography variation on as-prepared samples. The rectangular area shows the EBSD scan region~($1030 \ \rm{px} \times 1030 \ \rm{px}$) 
			\item The unprocessed EBSD inverse pole figure~(IPF) $z$ map, with $z$ axis orthogonal to the surface;
			\item \hkl {001}, \hkl{110}, \hkl{111} pole figures showing the complete texture information.
		\end{enumerate*}}
		\label{fig:pre_sem}
	\end{figure}
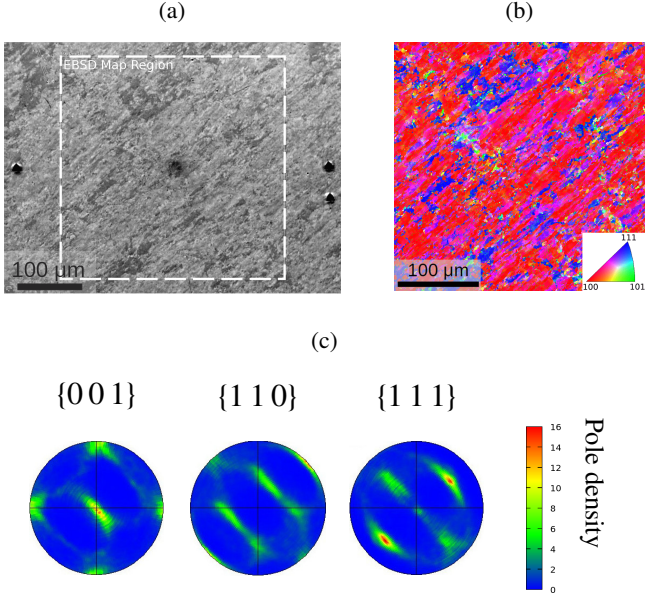

The SEM used in this work was the TESCAN Mira3 XMH, which uses a Schottky field emission gun and a NordlysNano EBSD detector. For EBSD mapping~(Fig.~\ref{fig:pre_sem}) samples were placed on a~$70${\textdegree} pre-tilted specimen holder and oriented so that the indents were aligned vertically in the view field~(as in Fig.~\ref{fig:fiducial}). For EBSD, the scan region was centred exactly between the two larger indents, and its size set to~$350 \ \mu\textrm{m} \times 350 \ \mu\textrm{m}$~(pixel size~$\leq 0.33 \ \mu\textrm{m}$). Due to the tilt and long scanning time~($2$~hours), samples experienced small drift effects, which were corrected automatically by using the Aztec drift correction. Kikuchi pattern acquisition was conducted according to the ISO 13067:2011 standard. The percentage of unresolved pixels was less than~$10$~\%, which allowed effective noise elimination using HKL CHANNEL5 post-processing software~(six nearest neighbour averaging). For grain boundary detection, the critical angle was set to~$10${\textdegree} and minimum grain size to~$10$ pixels. The average grain size in all samples was calculated to be~$(3.0 \pm 0.5) \ \mu\rm{m}$.

\subsubsection{Raman spectroscopy and CLSM}
\label{sect:raman_settings}

Raman spectroscopy is a powerful tool to study tungsten oxides. It has been used recently to characterise structural changes in tunable electrochromic devices based on nanocrystalline WO$_3$ as a function of applied voltage~\cite{MicroRaman1}. In another application, thin~(up to~$200$~nm) oxide layers formed on a tungsten metal before and after exposure to deuterium or helium plasma have been characterised with Raman spectroscopy~\cite{RamanWO,RamanWO2}. CLSM, although its resolution is ultimately limited by the diffraction limit, has been successfully used previously to examine small topography variations of less than~$0.5 \ \mu\rm{m}$ in oxidised and sputtered tungsten in~\cite{Preferential1,Sputtering2}. 

In this work, the WITec~alpha300~AR confocal Raman imaging microscope was used to acquire Raman spectra in the raster mode and perform large area scans~(Raman and CLSM maps). The TrueSurface feature enabled dynamic focusing of the confocal laser during scanning. For the tested samples, the initial surface roughness, defined here by the inter-quartile range~(IQR) of the depth coordinate variation gathered from the $350 \ \mu\textrm{m} \times 350 \ \mu\textrm{m}$~(pixel size $1.4 \ \mu\rm{m}$) region, was $0.176 \pm 0.016~\mu$m. By accumulating a large number of pixels ($\ge 10^{6}$), the error in the IQR surface topography measurement is reduced by a factor of $1000$ and therefore, the resolution for this oversampled dataset is $0.5$~nm. Post-oxidation confocal mapping in combination with the Raman spectroscopy were carried out on $75 \ \mu\textrm{m} \times 75 \ \mu\textrm{m}$~(pixel size~$0.05 \ \mu\rm{m}$, see Fig.~\ref{fig:fiducial}) areas. Single point Raman spectra were acquired with the conventional CCD camera in the low-noise high-intensity detection mode by averaging over~$100$~consecutive accumulations, each accumulation taking up to~$0.5$~s. Large area scans~($75 \ \mu\textrm{m} \times 75 \ \mu\textrm{m}$) were performed with an electron multiplying charge coupled device~(EMCCD) camera in the low intensity mode, allowing fast acquisition~($1 \times 10^{-3}$~s per pixel). All measurements were carried out using a green laser~($\lambda = 531.95 \ \textrm{nm}$), a~$1800 \ \textrm{g}/\textrm{mm}$ grating and a Zeiss EC~\enquote{Epiplan-Neofluar} $100 \times$ objective with a numerical aperture (NA) of $0.90$. Under these conditions, the diameter of the laser spot was~$0.72 \ \mu\rm{m}$, which required keeping the laser power below~$3 \ \textrm{mW}$ to avoid permanent damage to the oxidised material. The pixel size for EMCCD scans~$0.05 \ \mu\rm{m}$ is smaller than the optical resolution for this objective~($\approx 0.36 \ \mu\rm{m}$); this was designed on purpose as to eliminate the statistical error due to EMCCD read-out~(thermal and read noise). 

\subsection{Data processing methods}
\label{sect:data_processing}

Raman spectroscopy is based on the inelastic scattering of photons. In this process, the material under test~(MUT) is polarised by an incoming photon of any frequency, generating a virtual excited energy state. If a change in vibrational state of the MUT induces a change in polarisation, then the energy of the re-radiated photon is shifted. This shift in energy can be measured and is known as Raman scattering. However, this process is very inefficient and non-resonant. On the other hand, fluorescence caused by the excitation of MUT electronic states, is a highly efficient, resonant process. Therefore, accumulated Raman spectra suffer from a superimposed baseline signal caused by fluorescence and for low laser powers, are sensitive to noise. Both smoothing to remove noise and a baseline subtraction are required to access the underlying Raman signal, which contains the material bond-specific characteristic vibrational energies. 

\subsubsection{Smoothing based on LOESS filter with BIC statistic}

In order to remove noise from the baseline Raman signal, a locally estimated scatterplot smoothing~(LOESS) filter~\cite{kiselev2016design} is used. This LOESS algorithm has one parameter, the window size, which sets the smoothing region. For each window, a second order polynomial is fitted via a standard least squares regression with no weighting term. The ideal number of windows was calculated by using the Bayesian Information Criterion (BIC) \cite{InformationCriterion}. In this approach, the sum of least squares is penalised by an additional term which is weighted by the parameter,~$k$. The BIC statistic is then defined as: 
\begin{equation}
\centering
\label{eqn:BIC statistic}
\mathrm{BIC} = n\log{\sigma_{e}^{2}} + k\log{n}
\end{equation}
where $n$ is the number of measured points, $\sigma_{e}^{2}$ is the mean square error and $k$ is the number of parameters.

\begin{figure*}[!ht]
	\centering
	\captionsetup{singlelinecheck=off,skip=0pt}
			\begin{tikzpicture}
			\begin{groupplot}
			[group style={
				group name=smoothing,
				group size=1 by 2,
				xlabels at=edge bottom,
				xticklabels at=edge bottom,
				vertical sep=10pt,			
			},
			xlabel={LOESS parameter},
			xmin=0,
			xmax=100,
			xtick align=outside,
			every y tick scale label/.style={at={(yticklabel cs:0.5)}, anchor = south, rotate = 90,},
			xtick pos=left,
			ytick pos=left,
			scale only axis,
			minor tick num=1,
			grid=both,
			major grid style={dotted,black!75!white,very thin},
			minor grid style={dotted,black!75!white,very thin},
			width=0.175\textwidth,
			y tick label style={
				/pgf/number format/.cd,
				fixed,
				fixed zerofill,
				precision=2,
				/tikz/.cd
			},
			]
			\nextgroupplot[ytick scale label code/.code={$\chi^2$~\pgfmathparse{int(#1)}$(\times 10^{\pgfmathresult})$},				title={(\textbf{a})}]
			\addplot[color=red,mark=square*,very thick,mark options={scale=0.5,fill=red,draw=black,very thin}]
			table[x=Loess,y=Chi**2] {Data/Fig3a-b_numerical-data_chi-squared_BIC.txt};
			\coordinate (BOTTOM) at (rel axis cs:1.75,-0.65);		
			\nextgroupplot[ytick scale label code/.code={BIC~\pgfmathparse{int(#1)}$(\times 10^{\pgfmathresult})$}]
			\addplot[color=blue,mark=*,very thick,mark options={scale=0.5,fill=blue,draw=black,very thin}]
			table[x=Loess,y=BIC] {Data/Fig3a-b_numerical-data_chi-squared_BIC.txt};
			\end{groupplot}
			\begin{axis}[
			xlabel={rel. wavenumber~(cm$^{-1}$)},
			ylabel={CCD counts},
			xmin=200,
			xmax=1200,,		
			title={(\textbf{b})},
			legend style={ at={(rel axis cs: 0.5, -0.5)}, anchor=south,legend columns=3,fill=none,align=center,thin},
			xtick pos=left,
			ytick pos=left,
			at={(BOTTOM),anchor=south west},
			width=0.35\textwidth,
			]
			\addplot[color=blue!35!white,mark=\empty,thick] [mark options={scale=1.25}]
			table[x=Raw_Wavenumber,y=Raw_CCD] {Data/Fig3c_numerical-data_Smoothed_Baseline_Raw.txt};
			\addlegendentry{Raw}
			\addplot[color=red,mark=\empty,very thick]
			table[x=Rel_wavenumber,y=Smoothed_CCD] {Data/Fig3c_numerical-data_Smoothed_Baseline_Raw.txt};
			\addlegendentry{Smoothed}
			\addplot[color=green!65!black,mark=\empty,densely dashed]
			table[x=Rel_wavenumber,y=Baseline_CCD] {Data/Fig3c_numerical-data_Smoothed_Baseline_Raw.txt};
			\addlegendentry{Baseline}
			\end{axis}
			\end{tikzpicture}		
		%
	\caption[foo_bar]{LOESS filter smoothing of W oxide Raman spectrum: \begin{enumerate*}[label=(\alph*)]
			\item Mean squared error (MSE) and  Bayesian information criterion (BIC) variation with LOESS window number;
			\item Resulting spectrum after applying a LOESS filter with window size corresponding to the local minima of BIC statistic. The baseline is calculated using the asymmetric least squares~(AsLS) fitting procedure.
	\end{enumerate*}}
	\label{fig:Raman loess smoothing}
\end{figure*}
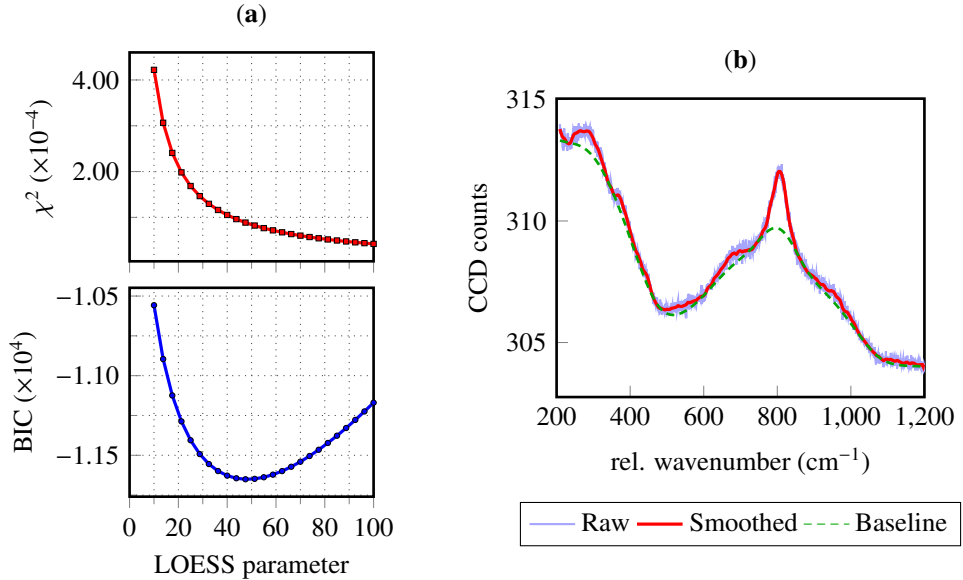

The number of windows, $m$ is directly proportional to this penalisation factor~$k$. The coefficient of proportionality is equal to four, arising from the three variables ($\theta_{0}, \theta_{1}, \theta_{2})$ required for second order polynomial fitting and the additional variance term. An example of this fitting is shown in Fig.~\ref{fig:Raman loess smoothing}.

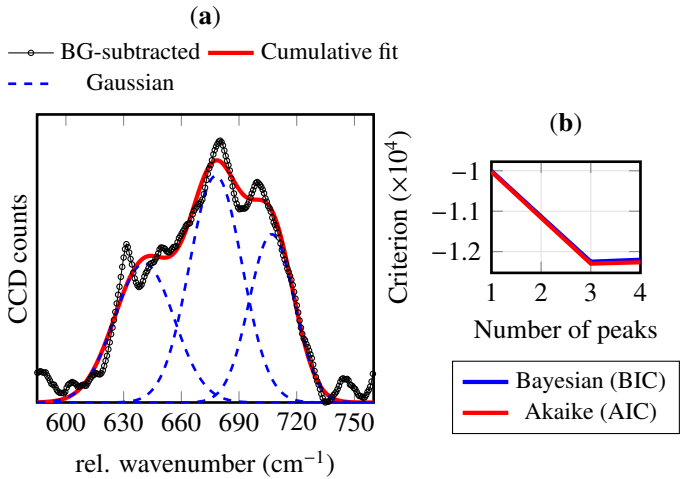
\begin{figure}[!ht]
	\centering
	\captionsetup{singlelinecheck=off,skip=0pt}
	\centering
		%
			\begin{tikzpicture}
			\begin{axis}[
			scale only axis,
			xlabel={rel. wavenumber~(cm$^{-1}$)},
			ylabel={CCD counts},
			xmin=585,
			xmax=760,
			ymin=0,
			legend pos=north west,
			xtick={600,630,660,...,750},
			width=0.5\columnwidth,
			legend style={ at={(rel axis cs: 0.5, 1.03)}, anchor=south,legend columns=2,fill=none,draw=none,align=center,font=\small},
			title={(\textbf{a})},
			title style={yshift={5.0ex}},
			ytick=\empty,
			]
			\addplot[mark=o,thin] [mark options={scale=0.5}]
			table[x=Rel_wavenumber,y=Baseline_subtracted_CCD] {Data/Fig4a_Baseline-subtracted-CCD_rel-wavenumber.txt};
			\addlegendentry{BG-subtracted}
			\addplot[domain=585:760,samples=200,mark=\empty, red, ultra thick] { f(x,20.5984929,13.7342887,677.922988)  
				+ f(x,13.0144338,11.6888754,707.028610)
				+ f(x,14.2930314,15.5364582,640.706976)};	
			\addlegendentry{Cumulative fit}
			\addplot[domain=585:760,samples=200,mark=\empty, dashed, blue] { f(x,14.2930314,15.5364582,640.706976)  };
			\addlegendentry{Gaussian}
			\addplot[domain=585:760,samples=200,mark=\empty, dashed, blue] { f(x,13.0144338,11.6888754,707.028610)  };
			\addplot[domain=585:760,samples=200,mark=\empty, dashed, blue] { f(x,20.5984929,13.7342887,677.922988)  };
			\coordinate (POS1) at (rel axis cs:1.35,0.45);		
			\end{axis}
			\begin{axis}[	
			xlabel={Number of peaks},
			xmin=1,
			xmax=4,
			legend pos=north east,
			grid=both,
			major grid style={line width=.2pt,draw=gray!25},
			xtick={1,2,...,5},
			ytick scale label code/.code={Criterion~\pgfmathparse{int(#1)}$(\times 10^{\pgfmathresult})$},
			every y tick scale label/.style={at={(yticklabel cs:0.5)}, anchor = south, rotate = 90},
			width=0.4\columnwidth,
			at={(POS1),anchor=west},
			title={(\textbf{b})},
			legend style={ at={(rel axis cs: 0.5, -1.45)}, anchor=south,legend columns=1,fill=none,align=center, thin,font=\small},
			]
			\addplot[mark=\empty,ultra thick,color=blue] 
			table[x=Number_Peaks_BIC,y=BIC_statistic] {Data/Fig4b_AIC_BIC_peaks.txt};
			\addlegendentry{Bayesian~(BIC)}
			\addplot[mark=\empty,ultra thick,color=red] 
			table[x=Number_Peaks_AIC,y=AIC_statistic] {Data/Fig4b_AIC_BIC_peaks.txt};
			\addlegendentry{Akaike~(AIC)}
			
			\end{axis}
		\end{tikzpicture}
	\caption[foo_bar]{Gaussian deconvolution based on the Levenberg-Marquardt damped least-squares~(DLS) method on model data~(smoothed, baseline subtracted):\begin{enumerate*}[label=(\alph*)]
	\item Result of peak deconvolution in a narrow spectral range;
	\item Information criterion plotted against the number of peaks used as a fitting variable.
	\end{enumerate*}
}
	\label{fig:deconvolution_example}
\end{figure}

\subsubsection{Baseline subtraction based on asymmetric least squares smoothing}

In order to remove the superimposed baseline from the Raman signal, an asymmetric least squares~(AsLS) approach is taken~\cite{Baseline1}. This fitting procedure depends on two parameters: $\lambda$ and $p$. $\lambda$ is the curvature penalisation term, and $p$ determines the asymmetry in the least squares iterations:

\begin{equation}
\centering
\label{eqn:Asymmetric least squares}
S = \sum_{i}^{n}{w_{i}}(y_{i}-z_{i})^{2} + \lambda \sum_{i}^{n}(\Delta^{2}{z_{i}})
\end{equation}
where $w_{i} = p$, if $y_{i} > z_{i},$ else $w_{i} = (1-p)$. Typically $\lambda = 10^{5}$~--~$10^{7}$ and $p=0.001$~--~$0.05$.

The AsLS is encoded in the weighting term~$w_{i}$. If the calculated baseline intensity~$z_{i}$ is higher than the measured signal, then the weighting term is equal to $(1-p)$ and is large. A set of linear equations is solved iteratively in reference to~\cite{eilers2005baseline}. Fig.~\ref{fig:Raman loess smoothing} illustrates the baseline fitting procedure for a model spectrum.

\subsubsection{Gaussian peak deconvolution}

A Gaussian peak deconvolution software package based on \lstinline{python}'s \lstinline{lmfit} module was used to deconvolute the peak positions for the baseline subtracted W oxide Raman signal. The Levenberg-Marquardt algorithm, a damped least-squares~(DLS) method, was used in combination with a BIC statistic to determine the most likely number of Gaussian peaks and their parameters~(peak position, FWHM and amplitude) which form up the tested spectra. Gaussian functions were chosen to fit the underlying spectra as is typical for solids with $\tau_{a} \gg \tau_{c}$, where $\tau_{a}$ is the lifetime of the vibrationally excited bond and $\tau_{c}$ is the de-phasing lifetime. The small amplitude of lattice vibrations in tungsten at low temperatures ensures that the de-phasing lifetime is small~\cite{Bradley2015RamanLineshape}. Fig.~\ref{fig:deconvolution_example} illustrates the deconvolution process on an example W oxide Raman spectrum.

\subsection{Oxidation procedure}
\label{sect:oxidation}

A controlled high-temperature oxidation experiment in tungsten~($T=800$--$1200$~{\textdegree}C typical to the monoblock divertor during temperature transients) is challenging as it requires a high-vacuum system and sensitive oxygen sensors. Conversely, oxidation behaviour of tungsten at moderate temperatures~($T=400$--$600$~{\textdegree}C) will determine the performance of the first wall tungsten armour at the onset of LOCA~\cite{LOCA_Temperature}. For this study, a single temperature~$T=400$~{\textdegree}C was chosen. Oxidation was completely suppressed at characteristic times of the experiment under a vacuum~$p \approx 10^{-4}$~bar. Increasing the chamber pressure to~$p_{\rm{Ar}-\rm{O}_2}=0.1$--$1.0$ for~$t=20$~min allowed the growth of a thin oxide layer in a controlled manner. Oxidation was negligible during the chamber filling/evacuation transients, resulting in a nearly-static, controlled atmosphere suitable for discrete oxidation. The test temperature $T=400$~{\textdegree}C is close to the temperature range considered by \citet{Preferential1} in their recent study, which facilitates comparison.

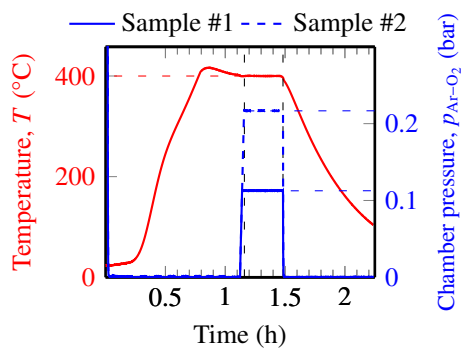
\begin{figure}
	\centering
	\captionsetup{singlelinecheck=off}	
	\begin{tikzpicture}
		\pgfplotsset{
			xmin=0, xmax=2.25,
			xlabel={Time~(h)},
			xtick={0.5,1,...,2},
			minor x tick num=4,
			ymin=0,		
			axis y line*=left,
			scale only axis,
			width=0.4\columnwidth,
		}	
		\def\xmin{\pgfkeysvalueof{/pgfplots/xmin}}
		\def\xmax{\pgfkeysvalueof{/pgfplots/xmax}}
		\def\ymin{\pgfkeysvalueof{/pgfplots/ymin}}
		\def\ymax{\pgfkeysvalueof{/pgfplots/ymax}}		
		\begin{axis}[
		ylabel={Temperature,~$T$~({\textdegree}C)},
		y tick label style={red},
		y label style={red},				
		]
		\addplot[color=red,mark=\empty, thick] 
		table[x expr=\thisrowno{0}/60/60,y=Temperature] {Data/TemperaturePressureVariation.dat}
		coordinate [pos=0.59] (A);		
		\draw [loosely dashed,red,thin]let \p1 = (A) in (A) -- (\xmin,\y1);			
		\draw [loosely dashed,black,thin] (axis cs:1.16,\ymin) -- (1.16,\ymax);					
		\draw [loosely dashed,black,thin] (axis cs:1.4825,\ymin) -- (1.4825,\ymax);							
		\end{axis}
		\begin{axis}[xlabel={},
		ylabel={Chamber pressure,~$p_{\rm{Ar} - \rm{O}_2}$~(bar)},
		y tick label style={blue},	
		y label style={blue,font=\small},
		x label style={font=\small},	
		axis y line*=right,
		ymin=0,
		ymax=0.3,
		ytick={0,0.1,...,0.3}, 
		legend style={at={(rel axis cs: 0.5, 1.1)}, anchor=south,legend columns=2,fill=none,draw=none,anchor=center,align=center},
		]
		\addplot[color=blue,mark=\empty, thick] 
		table[x expr=\thisrowno{2}/60/60,y expr=\thisrowno{3}+0.9978] {Data/TemperaturePressureVariation.dat}
		coordinate [pos=0.65] (B);
		\draw [loosely dashed,blue,thin]let \p2 = (B) in (B) -- (\xmax,\y2);
		\addlegendentry{Sample~\#1};
		\addplot[color=blue,mark=\empty, thick,dashed] 
		table[x expr=\thisrowno{4}/60/60,y expr=\thisrowno{5}+1.01] {Data/TemperaturePressureVariation.dat}
		coordinate [pos=0.65] (C);
		\draw [loosely dashed,blue,thin]let \p3 = (C) in (C) -- (\xmax,\y3);	
		\addlegendentry{Sample~\#2};
		\end{axis}	
		\coordinate (cc) at (rel axis cs:0.5,1.03);		
		\end{tikzpicture}
	\caption[foo_bar]{Measured variation of temperature~$T$ and chamber pressure~$p_{\rm{Ar}-\rm{O}_2}$ against time in experiments to study the onset of oxidation at~$T=400$~{\textdegree}C}
	\label{fig:sta}
\end{figure}

Tungsten samples~(Table~\ref{tbl:samples}) were placed in Al$_2$O$_3$ crucibles with a~$0.12$~ml capacity, polished side facing upwards. No lid was used to cover the samples. The crucibles were then placed inside the heating stage of a Linseis STA PT1600 simultaneous thermal analyser. The chamber was then evacuated and samples were heated up to~$400$~{\textdegree}C. The initial heating rate~$20$~{\textdegree}C/min was changed half-way through the heating process to~$10$~{\textdegree}C/min~(Fig.~\ref{fig:sta}) to decrease thermal inertia. After thermal equilibration, the instrument chamber was filled with Ar--20~\%~O$_2$ at a flow rate~$38$~l/h. Isothermal oxidation segments were designed differently to model:

\begin{table}[]
	\caption{A summary table for samples oxidised at~$T=400$~{\textdegree}C}
	\label{tbl:samples}
	\centering
	\begin{tabular}{rrrr}
		\hline
		\multicolumn{1}{c}{\#} & \begin{tabular}[c]{@{}c@{}}Chamber pressure, \\ $p_{\rm{Ar}-\rm{O}_2}$  (bar)\end{tabular} & \begin{tabular}[c]{@{}c@{}} Oxidation \\ time, $t$ \end{tabular} & Dominant colour \\ \hline
		1 & 0.12 & \multirowcell{3}{\rotatebox[origin=c]{90}{20~min}} & Pale Yellow \\ 
		2 & 0.22 & & Yellow \\
		3 & \multirow{5}{*}{\rotatebox[origin=c]{90}{Atmospheric}} &  & Yellow + Brown \\
		4 &  & 1~h & Dark Purple \\
		5 &  & 5~h & Sky Blue \\
		6 &  & 10~h & Purple \\
		7 &  & 72~h & Blue \\
		\hline
	\end{tabular}%
\end{table}

\begin{enumerate}[label=(\alph*)]
	\item \textit{Initial oxidation stages}. An under-pressure~($p_{\rm{Ar}-\rm{O}_2} < 1.0 \ \rm{bar}$) was created in the chamber for a short period of time~($t=20$~min, Fig.~\ref{fig:sta}), which is still much longer than the transient period when the gas was flowing in or out;  
	\item \textit{Progression of oxidation}. Oxidation runs were conducted at atmospheric pressure~($p_{\rm{Ar}-\rm{O}_2} = 1.0 \ \rm{bar}$). The oxidation time was varied from~$t=20$~min to~$t=72$~h. 
\end{enumerate}

\noindent The chamber was then evacuated and the furnace cooled down to room temperature at~$20$~{\textdegree}C/min. 

\section{Results and Discussion}

\subsection{Summary of reference spectra}

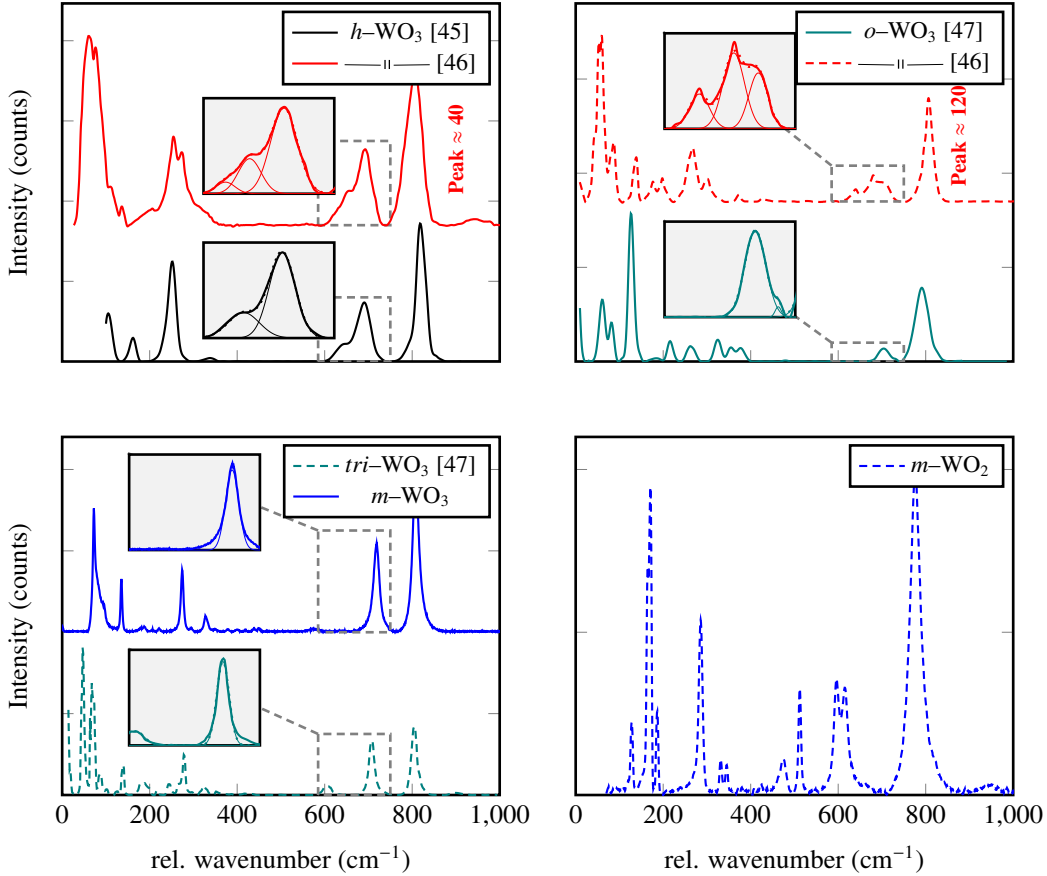
\begin{figure*}[!ht]
	\centering
	
		\begin{tikzpicture}
		\begin{groupplot}
		[group style={
			group name=references,
			group size=2 by 2,
			xlabels at=edge bottom,
			xticklabels at=edge bottom,
			ylabels at=edge left,
			yticklabels at=edge left,
		},
		xlabel={rel. wavenumber (cm$^{-1}$)},
		ylabel={Intensity (counts)},
		xmin=0,
		xmax=1000,
		ymin=0,
		legend style={font=\small},
		xtick pos=left,
		yticklabels={,,},
		width=0.4\textwidth,		
		]
		\nextgroupplot
		\addplot[color=black,mark=\empty,thick]
		table {Data/Reference_Spectra/Hexagonal_Daniel.txt};
		\coordinate (insetPosition1) at (axis cs:320,0.75);
		\coordinate (insetPosition2) at (axis cs:320,1.65);
		\coordinate (labelPosition1) at (axis cs:900,1.325);
		\node[rotate=90,red,ultra thick] at (labelPosition1) {\footnotesize{\textbf{Peak $\approx$ 40}}};
		\draw [gray, densely dashed] (axis cs:585,0.0) rectangle (axis cs:750,0.4);	
		\draw [gray, densely dashed] (axis cs:585,0.85) rectangle (axis cs:750,1.375);	
		\addlegendentry{ $h$--WO$_3$~\cite{Daniel1987} }	
		\addplot[color=red,mark=\empty,thick]
		table [y expr=\thisrowno{1}+0.85] {Data/Reference_Spectra/Hexagonal_Pecquenard.txt};
		\addlegendentry{ \dittotikz~\cite{Pecquenard1998} };	
		\nextgroupplot
		\addplot[color=teal,mark=\empty,thick]
		table {Data/Reference_Spectra/Orthorhombic_Gabrusenoks.txt};
		\addlegendentry{ $o$--WO$_3$~\cite{Gabrusenoks2001} };
		\addplot[densely dashed,color=red,mark=\empty,thick]
		table [y expr=\thisrowno{1}+0.85] {Data/Reference_Spectra/Orthorhombic_2.txt}; 
		\addlegendentry{ \dittotikz~\cite{Pecquenard1998} };	
		\coordinate (insetPosition3) at (axis cs:200,0.75);
		\coordinate (insetPosition4) at (axis cs:200,1.75);
		\coordinate (lineStart1) at (axis cs:585,0.1);
		\coordinate (lineStart2) at (axis cs:585,1.04);
		\coordinate (labelPosition2) at (axis cs:875,1.2);		
		\draw [gray, densely dashed] (axis cs:585,0.0) rectangle (axis cs:750,0.1);	
		\draw [gray, densely dashed] (axis cs:585,0.85) rectangle (axis cs:750,1.04);
		\draw [gray, densely dashed] (lineStart1) -- (insetPosition3);
		\draw [gray, densely dashed] (lineStart2) -- (insetPosition4);
		\node[rotate=90,red,ultra thick] at (labelPosition2) {\footnotesize{\textbf{Peak $\approx$ 120}}};
		\nextgroupplot
		\addplot[color=teal,mark=\empty,thick,densely dashed]
		table[y expr=\thisrowno{1}*2.0]{Data/Reference_Spectra/Triclinic.txt};	
		\addlegendentry{ $tri$--WO$_3$~\cite{Gabrusenoks2001} };		
		\addplot[color=blue,mark=\empty,thick]
		table[y expr=\thisrowno{1}+1] {Data/Reference_Spectra/m_WO3.txt};
		\addlegendentry{ $m$--WO$_3$ };	
		\coordinate (insetPosition5) at (axis cs:150,0.9);
		\coordinate (insetPosition6) at (axis cs:150,2.1);
		\coordinate (lineStart3) at (axis cs:585,0.375);
		\coordinate (lineStart4) at (axis cs:585,1.625);
		\draw [gray, densely dashed] (axis cs:585,0.0) rectangle (axis cs:750,0.375);	
		\draw [gray, densely dashed] (axis cs:585,1.0) rectangle (axis cs:750,1.625);
		\draw [gray, densely dashed] (lineStart3) -- (insetPosition5);
		\draw [gray, densely dashed] (lineStart4) -- (insetPosition6);	
		\nextgroupplot
		\addplot[color=blue,mark=\empty,thick,densely dashed]
		table {Data/Reference_Spectra/WO2.txt};			
		\addlegendentry{ $m$--WO$_2$ };	
		\end{groupplot}
		\begin{axis}[
		at={(insetPosition1)},anchor={outer north west},footnotesize,width=0.18\textwidth,xmin=600,xmax=750,ymin=0,ytick=\empty,xtick=\empty,axis background/.style={fill=gray!10}]
		\addplot[color=black,mark=\empty,thick]
		table {Data/Reference_Spectra/Hexagonal_Daniel.txt};
		\addplot[domain=585:760,samples=200,mark=\empty, black,very thin] 
		{ g(x, 13.65923, 34.928, 690.37803) };	
		\addplot[domain=585:760,samples=200,mark=\empty, black,very thin] 
		{ g(x, 4.89948, 42.1538, 646.12185) };	
		\addplot[domain=585:760,samples=200,mark=\empty, black,thick, dotted] 
		{ g(x, 4.89948, 42.1538, 646.12185) + g(x, 13.65923, 34.928, 690.37803) };	
		\end{axis}
		
		\begin{axis}[
		at={(insetPosition2)},anchor={outer north west},footnotesize,width=0.18\textwidth,xmin=600,xmax=750,ymin=0,ytick=\empty,xtick=\empty,axis background/.style={fill=gray!10}]
		\addplot[color=red,mark=\empty,thick]
		table {Data/Reference_Spectra/Hexagonal_Pecquenard.txt};
		\addplot[domain=585:760,samples=200,mark=\empty, red,very thin] 
		{ g(x, 17.9784, 36.25163, 692.28489) };	
		\addplot[domain=585:760,samples=200,mark=\empty, red,very thin] 
		{ g(x, 6.28433, 30.96723, 652.89564) };	
		\addplot[domain=585:760,samples=200,mark=\empty, red,very thin] 
		{ g(x, 1.72535, 25.68356, 625.22896) };
		\addplot[domain=585:760,samples=200,mark=\empty, red, thick, dotted] 
		{ g(x, 1.72535, 25.68356, 625.22896) + g(x, 6.28433, 30.96723, 652.89564) + g(x, 17.9784, 36.25163, 692.28489) };	
		\end{axis}
		\begin{axis}[
		at={(insetPosition3)},anchor={outer north west},footnotesize,width=0.18\textwidth,xmin=600,xmax=750,ymin=0,ytick=\empty,xtick=\empty,axis background/.style={fill=gray!10}]
		\addplot[color=teal,mark=\empty,thick]
		table {Data/Reference_Spectra/Orthorhombic_Gabrusenoks.txt};
		\addplot[domain=585:760,samples=200,mark=\empty, teal,very thin] 
		{ g(x, 15.38543, 37.7734, 791.40826) };	
		\addplot[domain=585:760,samples=200,mark=\empty, teal,very thin] 
		{ g(x, 0.07295, 8.26831, 730.44706) };
		\addplot[domain=585:760,samples=200,mark=\empty, teal,very thin] 
		{ g(x, 2.08789, 28.23717, 704.11038) };			
		\addplot[domain=585:760,samples=200,mark=\empty, teal, thick, dotted] 
		{ g(x, 2.08789, 28.23717, 704.11038) + g(x, 0.07295, 8.26831, 730.44706) + g(x, 15.38543, 37.7734, 791.40826) };
		\end{axis}
		\begin{axis}[
		at={(insetPosition4)},anchor={outer north west},footnotesize,width=0.18\textwidth,xmin=600,xmax=750,ymin=0,ytick=\empty,xtick=\empty,axis background/.style={fill=gray!10}]
		\addplot[color=red,mark=\empty,thick]
		table {Data/Reference_Spectra/Orthorhombic_2.txt};
		\addplot[domain=585:760,samples=200,mark=\empty, red,very thin] 
		{ g(x, 2.40717, 25.33271, 707.52354) };	
		\addplot[domain=585:760,samples=200,mark=\empty, red,very thin] 
		{ g(x, 3.71865, 28.80391, 679.50362) };	
		\addplot[domain=585:760,samples=200,mark=\empty, red,very thin] 
		{ g(x, 1.60489, 27.14501, 639.91512) };
		\addplot[domain=585:760,samples=200,mark=\empty, red, thick, dotted] 
		{ g(x, 2.40717, 25.33271, 707.52354) + g(x, 3.71865, 28.80391, 679.50362) + g(x, 1.60489, 27.14501, 639.91512) };	
		\end{axis}
		\begin{axis}[
		at={(insetPosition5)},anchor={outer north west},footnotesize,width=0.18\textwidth,xmin=600,xmax=750,ymin=0,ytick=\empty,xtick=\empty,axis background/.style={fill=gray!10}]
		\addplot[color=teal,mark=\empty,thick]
		table {Data/Reference_Spectra/Triclinic.txt};
		\addplot[domain=585:760,samples=200,mark=\empty, teal,very thin] 
		{ g(x, 0.74454, 25.68373, 606.04179) };	
		\addplot[domain=585:760,samples=200,mark=\empty, teal,very thin] 
		{ g(x, 2.8674, 16.77568, 707.56147) };
		\addplot[domain=585:760,samples=200,mark=\empty, teal, thick, dotted] 
		{ g(x, 0.74454, 25.68373, 606.04179) + g(x, 2.8674, 16.77568, 707.56147)  };	
		\end{axis}
		\begin{axis}[
		at={(insetPosition6)},anchor={outer north west},footnotesize,width=0.18\textwidth,xmin=600,xmax=750,ymin=0,ytick=\empty,xtick=\empty,axis background/.style={fill=gray!10}]
		\addplot[color=blue,mark=\empty,thick]
		table {Data/Reference_Spectra/m_WO3.txt};
		\addplot[domain=585:760,samples=200,mark=\empty, blue,very thin] 
		{ f(x, 9.5, 7.58011168, 718.126449) };	
		\end{axis}
		\end{tikzpicture}
	\caption{Raman spectra for a variety of crystalline tungsten oxides: hexagonal WO$_3$~($h-$WO$_3$)~\cite{Gabrusenoks2001,Daniel1987}; orthorhombic WO$_3$~($o-$WO$_3$)~\cite{Gabrusenoks2001,Pecquenard1998}; monoclinic WO$_3$~($m-$WO$_3$); triclinic WO$_3$~($tri-$WO$_3$)~\cite{Gabrusenoks2001}; and monoclinic WO$_2$.}
	\label{fig:reference_spectra}
\end{figure*}

A well-developed Raman peak database for characterisation of tungsten oxides does not exist. Therefore, reference samples of WO$_3$~(Tungsten~(VI) oxide) and WO$_2$~(Tungsten~(IV) oxide) powders were purchased from Sigma-Aldrich and characterised with Raman scattering analysis. It was noticed that~WO$_2$ was unstable at a laser power above~$4 \ \textrm{mW}$~(surface power density~$\approx 1 \ \rm{MW}/{\rm{cm}^2}$). The instability also manifested itself as an irreversible change of colour from black to blue-violet. Different polymorphous states of~$\textrm{WO}_3$ which were not readily available for purchase were analysed by digitalising Raman spectra in published literature and subjecting them to the same baseline subtraction, smoothing and peak fitting as described in Sect.~\ref{sect:data_processing}. The resulting spectral shapes are provided in Fig.~\ref{fig:reference_spectra}, and~Table~\ref{tbl:big_reference_table} lists all the peak positions arranged so that the reader is easily able to match the occurrence of a specific peak to particular oxide phase.

\begin{table}[!ht]
	\caption{Results of peak deconvolution for all reference Raman spectra. Peak positions of major peaks are highlighted in bold. Grey rectangles outline the footprint peaks characteristic to a specific phase. Long dash indicates no data is available for this spectral position. Empty space means absence of peak.}
	\label{tbl:big_reference_table}
	\centering
	\resizebox{1.0\columnwidth}{!}{%
		\begin{tabular}{c:cc:cc:cc}
			\hline
			\multicolumn{7}{c}{Peak positions (rel. wavenumber (cm$^{-1}$))}  
			\\
			\multicolumn{1}{c}{\textbf{WO$_2$}} & \multicolumn{1}{c}{$tri$~-\textbf{WO$_3$}} & \multicolumn{1}{c}{\textbf{$m$~-WO$_3$}} & \multicolumn{2}{c}{$o$~-\textbf{WO$_3$}} & \multicolumn{2}{c}{$h$~-\textbf{WO$_3$}}    \\
			\textit{this work} & \cite{Gabrusenoks2001}    & \textit{this work}  & at~$650 \ \textrm{K}$~\cite{Gabrusenoks2001}  & at RT~\cite{Pecquenard1998} & \cite{Daniel1987} & \cite{Pecquenard1998} \\ \hline
			---                    & \multicolumn{1}{l}{\textbf{47}}                                &                                &                                                                                   & 43                                    &    ---                              &      \textbf{46}$^*$                               \\ ---
			&                                  &                                &                                                                                    & \textbf{52}                                    &      ---                             & \textbf{52}$^*$  \\ ---
			&                                  &                                & \textbf{61}                                                                                & \textbf{59}                                    &                        ---           &           \textbf{61}$^*$                         \\ \cellcolor[gray]{0.8} 75
			& \cellcolor[gray]{0.8}\textbf{70}                               & \cellcolor[gray]{0.8} \textbf{73}                             &                                                                                   &                                       &      ---                             &                                      \\
			& 87                               &                                & \textbf{82}                                                                                & 84                                    &                   ---                &           \textbf{80}$^*$                         \\ & 102 & & & & & \\
			&                                  &                                & 118                                                                               & 116                                      & 106                              & 116                                  \\
			\textbf{128}                        &  125                                & 123                            & \textbf{126}                                                                               &                                       &                                  &                                      \\
			& \textbf{139}                              & \textbf{135}                            &                                                                                   &      \textbf{134}$^*$                              &                                  & 137                                  \\
			&                                  &                                &                                                                                  &  \textbf{140}$^*$                                  &                                  &                                      \\
			\textbf{165}$^*$                        &                                  &                                &                                                                                   &                                       & \textbf{161}                              &                                      \\
			\textbf{171}$^*$                        &                                  &                                &                                                                                   & 176                                   &                                  &                                      \\
			186                        & 186                              & 184                            &  185                                                                                  &                                       &                                  &                                      \\
			&                                  &                                &                                                                                   & 199                                   &                                  & 200                                  \\
			&                 218                 & 216                            & 217                                                                               &                                       &                                  &                                      \\ 
			& 245                              &                                &                                                                                   &                                       & 247                              &                                      \\ 
			&                                  &                                &                                                                                   &                                       & \cellcolor[gray]{0.8} \textbf{253}                              & 
			\cellcolor[gray]{0.8}\textbf{254}$^*$                                  \\  
			&             263                     &                                & 263                                                                               & \textbf{265}                                   &                                  &                                      \\
			\textbf{285}                        & \textbf{278}                              & \textbf{274}                            &                                                                                   &                                       &                                  &                \textbf{275}$^*$                   \\
			&       296                           &                                &                                                                                   & 301                                   &                                  &                                      \\
			330                        & 324                              & 329                            & 325                                                                               &                                       & 336                              &                                      \\
			345                        &     352                             &                                & 354                                                                               &                                       &                                  &                                      \\ 
			&                                  &                                & \cellcolor[gray]{0.8}377                                                                               & \cellcolor[gray]{0.8}373                                   &                                  &                                      \\ 
			423        & 423                              &                                &                                                                                   & 429                                   &                                  &                                      \\
			\textbf{473}                        &                                  &                                &  479                                                                                 &                                       &                                  &                                      \\
			\textbf{512}                        &                                  &                                &                                                                                   &                                       &                                  &                                      \\
			&                                  & 574                            &                                                                                   &                                       &                                  &                                      \\
			\textbf{595}                        & \textbf{606}                              &                                &                                                                                   &                                       &                                  &                                      \\
			\textbf{615}                        &                                  &                                &                                                                                   &                                       &                                  & 625                                  \\ 660
			&                                  &                                &                                                                                   & 640                                   & 646                              & 653                                  \\ 696
			&                                  &                                &                                                                                   & \textbf{680}                                   & \cellcolor[gray]{0.8}\textbf{690}                              & \cellcolor[gray]{0.8}\textbf{692}                                  \\ 
			& \cellcolor[gray]{0.8}\textbf{708}                              & \cellcolor[gray]{0.8}\textbf{718}                            & \cellcolor[gray]{0.8}704                                                                               & \cellcolor[gray]{0.8}708                                   &                                  &                                      \\
			&                                  &                                & 730                                                                               &                                       &                                  &                                      \\
			\textbf{777}                        &                                  &                                &                                                                                   &                                       &                                  & 784                                  \\
			& \textbf{805}                              & \textbf{808}                            & \textbf{791}                                                                               & \textbf{807}                                   & \textbf{820}                              & \textbf{810}                                  \\
			&                                  &                                & 834                                                                               &                                       &                                  &                                      \\ 950
			&                                  &                                &                                                                                   &                                       &                                  & 946                              
			\\
			\hline
		\end{tabular}
	}
	\\
	{$^*$ Multiple overlaid peaks}
\end{table}%
Referring to~Fig.~\ref{fig:reference_spectra}, a few discrepancies between digitised literature spectral data are observed and this might be attributed to different instrumental settings. Hence, where possible, measurement conditions were kept to as described in the experimental procedure outlined in Sect.~\ref{sect:raman_settings}. The low-wavenumber region of Table~\ref{tbl:big_reference_table} shows that few peaks can be used as markers for a particular oxide phase -- especially when the signal/noise ratio is weak. In addition, bands below~$200 \ \mathrm{cm}^{-1}$ can be ascribed to lattice vibrations according to~\cite{MicroRaman}. Therefore, the~$600$--$1000$~$\mathrm{cm}^{-1}$ region of the spectra is used for phase analysis, as the spacing between characteristic peaks for different oxides phases seems to be sufficient for robust peak deconvolution. 

Raman spectra of all the~WO$_3$ polymorphous states considered in Table~\ref{tbl:big_reference_table} exhibit a dominant peak at~$805$--$820~\mathrm{cm^{-1}}$, which is attributed to the O--W stretching vibrations~\cite{Calcination1}. Hexagonal, orthorhombic, and monoclinic WO$_3$ phases are known to have very similar unit cell parameters~\cite{Hexagonal1,Orthorhombic1}, which are believed to correlate with the wavenumbers of the lattice stretching vibrations -- as in case of zeolites~\cite{Stretching1}. Hence, these O--W stretching bands are not expected to differ substantially; WO$_2$, having a different set of unit cell parameters, shows a peak at~$777$~cm$^{-1}$. On the other hand, the wavenumbers of bending vibrations~(following the nomenclature used in~\cite{Calcination1}) at~$640$--$718$~cm$^{-1}$ are very sensitive to the crystal structure and therefore all these oxides exhibit characteristic Raman peaks in this range. Hence they can be used primarily to distinguish between the different polymorphous states of WO$_3$.

Because non-oxidized tungsten does not show any significant Raman signal~(perhaps, only fluorescence, which contributes to the baseline), a sum-filter in the range $750-850~\mathrm{cm^{-1}}$ may be used as an indicator of the thickness of an oxide scale grown on W metal. Additional examination of the reference spectra shows that for the same experimental conditions~(i.e., laser power, spectrometer configuration, etc.), the Raman activity is weaker for some phases, which results in a lower intensity of the main peak, e.g.~$I\approx120$ for $o-$WO$_3$ versus~$I\approx40$ for~$h-$WO$_3$~(Fig.~\ref{fig:reference_spectra}).

\normalsize
\subsection{Initiation and growth: correlating EBSD and Raman scattering maps}
\label{sect:maps}

A chamber pressure~$p_{\mathrm{Ar-O}_{2}} \geq 0.1 \ \rm{bar}$ was required to produce sufficient Raman activity from the oxide. As such, the analysis starts from the least oxidised sample~(see Table \ref{tbl:samples}), tracks the progression of oxidation with increasing the chamber pressure up to atmospheric, and finally focuses on the samples exposed to oxygen for extended time periods (up to $72$~h).    

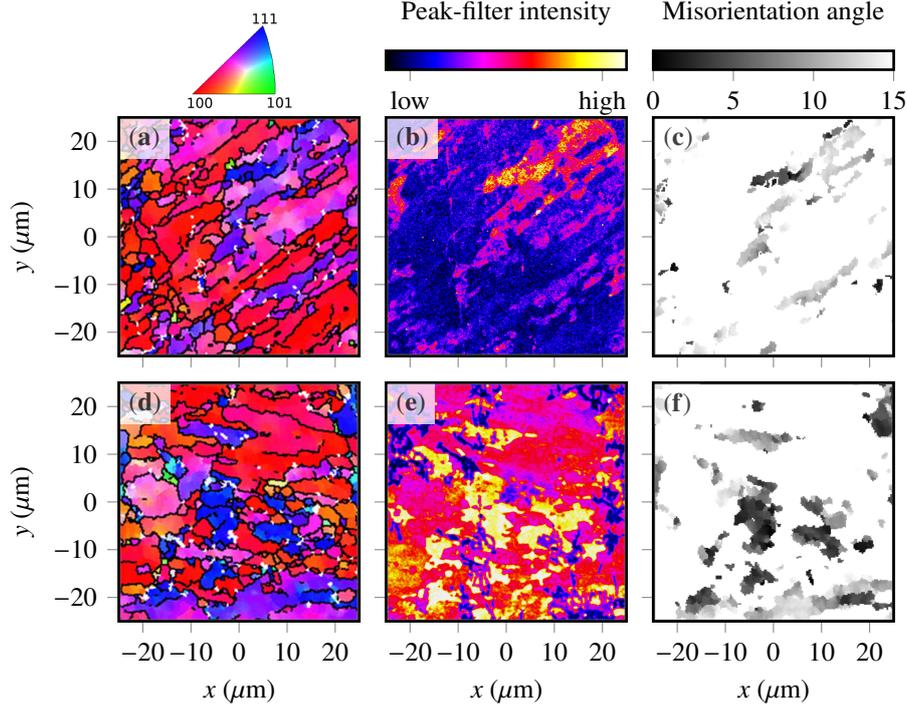
\begin{figure*}[!ht]
	\centering
	\captionsetup{singlelinecheck=off}
		\begin{tikzpicture}
		\begin{groupplot}
		[group style={
			group name=maps1,
			group size=3 by 2,
			xlabels at=edge bottom,
			xticklabels at=edge bottom,
			ylabels at=edge left,
			yticklabels at=edge left,
			vertical sep=10pt,	
			horizontal sep=10pt,	
		},
		xtick align=outside,
		ytick align=outside,
		xtick pos=left,
		ytick pos=left,
		scale only axis,
		xmin=-25,
		xmax=25,
		ymin=-25,
		ymax=25,
		xlabel={$x$~($\mu\rm{m}$)},
		ylabel={$y$~($\mu\rm{m}$)},
		xtick={-20,-10,...,20},
		ytick={-20,-10,...,20},
		axis on top=true,
		width=0.2\textwidth,
		]
		\nextgroupplot[axis equal image]
		\coordinate (ebsdLegend1) at (rel axis cs:0.5,1.03);	
		\addplot graphics[xmin=-25,ymin=-25,xmax=25,ymax=25] {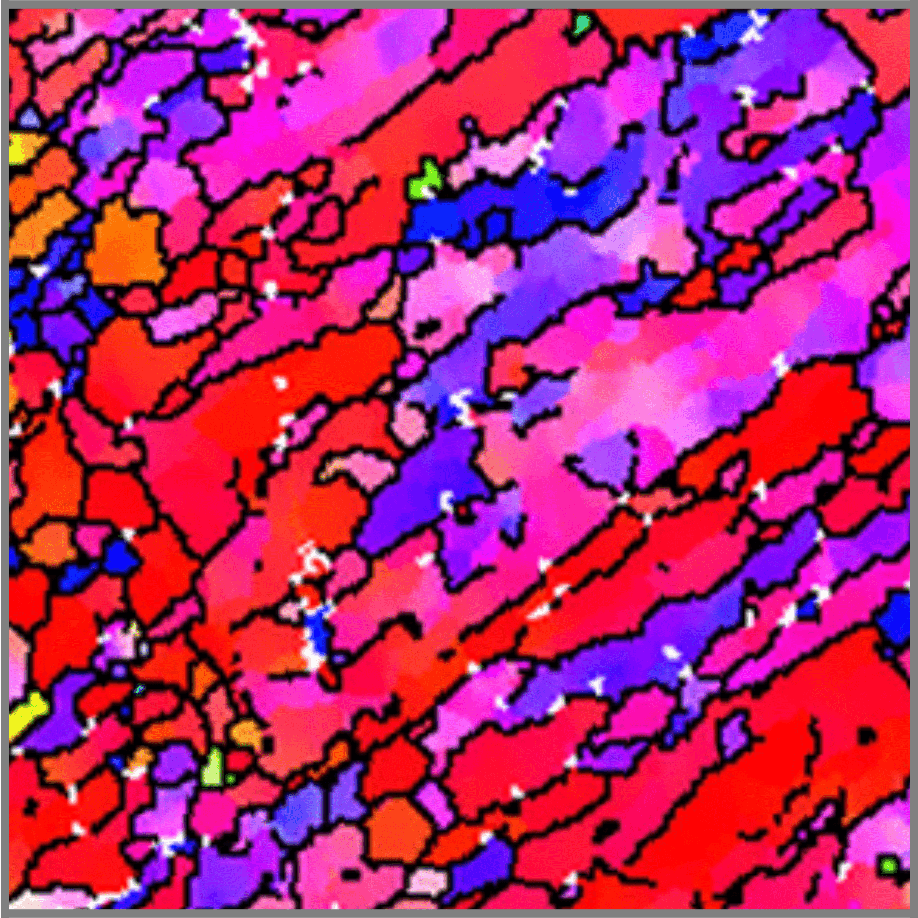};
		\node[anchor=north west,fill opacity=0.75,fill=white] at (rel axis cs:0,1) { {(\textbf{a})} };
		\nextgroupplot[
		axis equal image,
		colorbar horizontal,
		colormap={raman}{color(0cm)=(black) color(1cm)=(blue) color(2cm)=(magenta) color(3cm)=(red) color(4cm)=(yellow) color(5cm)=(white) },
		colorbar style={
			title=Peak-filter intensity,
			at={(0.5,1.2)},anchor=south,
			xticklabel pos=lower,
			xtick={0.1,0.9},
			xticklabels={low,high},
		},
		every colorbar/.append style={height=0.25cm}
		]
		\addplot graphics[xmin=-25,ymin=-25,xmax=25,ymax=25] {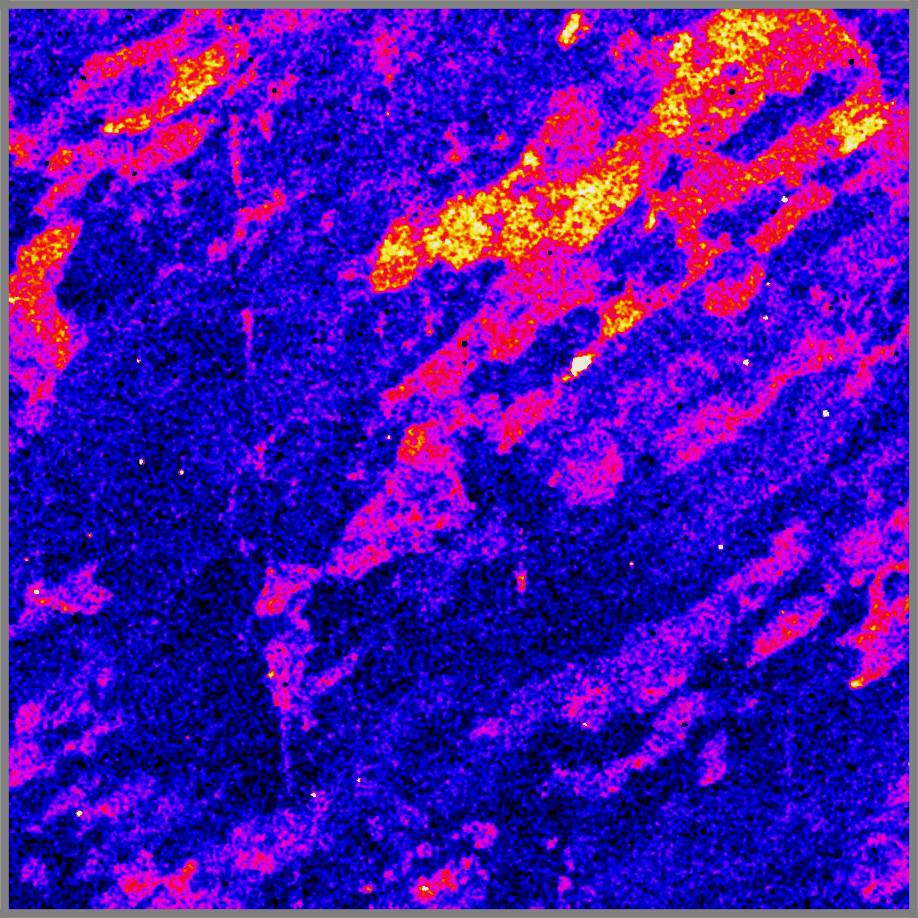};
		\node[anchor=north west,fill opacity=0.75,fill=white] at (rel axis cs:0,1) { {(\textbf{b})} };
		\nextgroupplot[
		axis equal image,
		colorbar horizontal,
		colormap/blackwhite,
		colorbar style={
			title=Misorientation angle,
			at={(0.5,1.2)},anchor=south,
			xticklabel pos=lower,
			point meta min=0.0,
			point meta max=15.0,
			xtick={0.0,5.0,...,15},
		},
		every colorbar/.append style={height=0.25cm}
		]
		\addplot graphics[xmin=-25,ymin=-25,xmax=25,ymax=25] {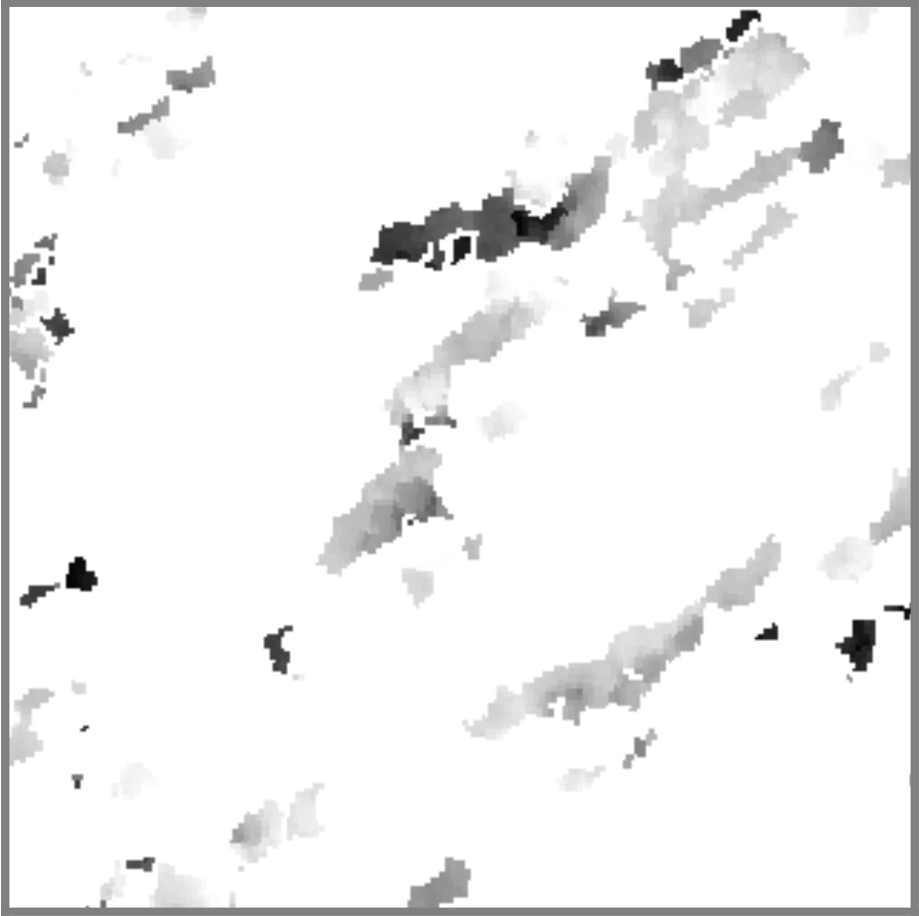};
		\node[anchor=north west,fill opacity=0.75,fill=white] at (rel axis cs:0,1) { {(\textbf{c})} };
		\nextgroupplot[
		axis equal image]
		\addplot graphics[xmin=-25,ymin=-25,xmax=25,ymax=25] {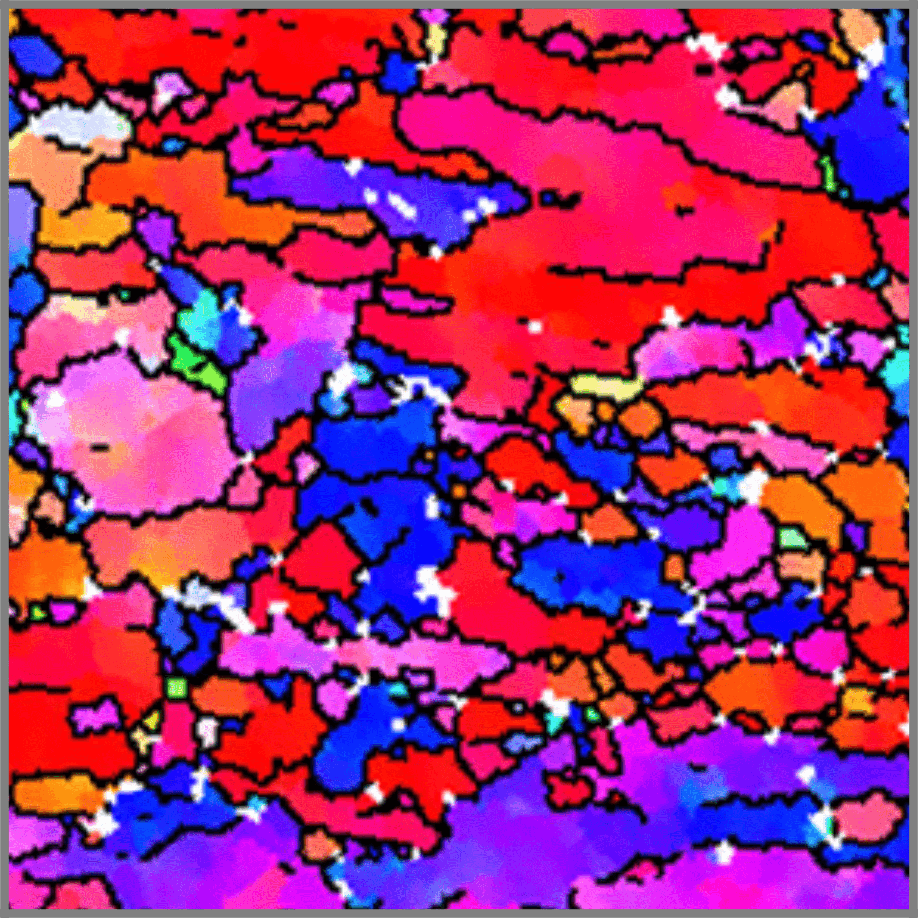};
		\node[anchor=north west,fill opacity=0.75,fill=white] at (rel axis cs:0,1) { {(\textbf{d})} };		
		\nextgroupplot[
		axis equal image]
		\addplot graphics[xmin=-25,ymin=-25,xmax=25,ymax=25] {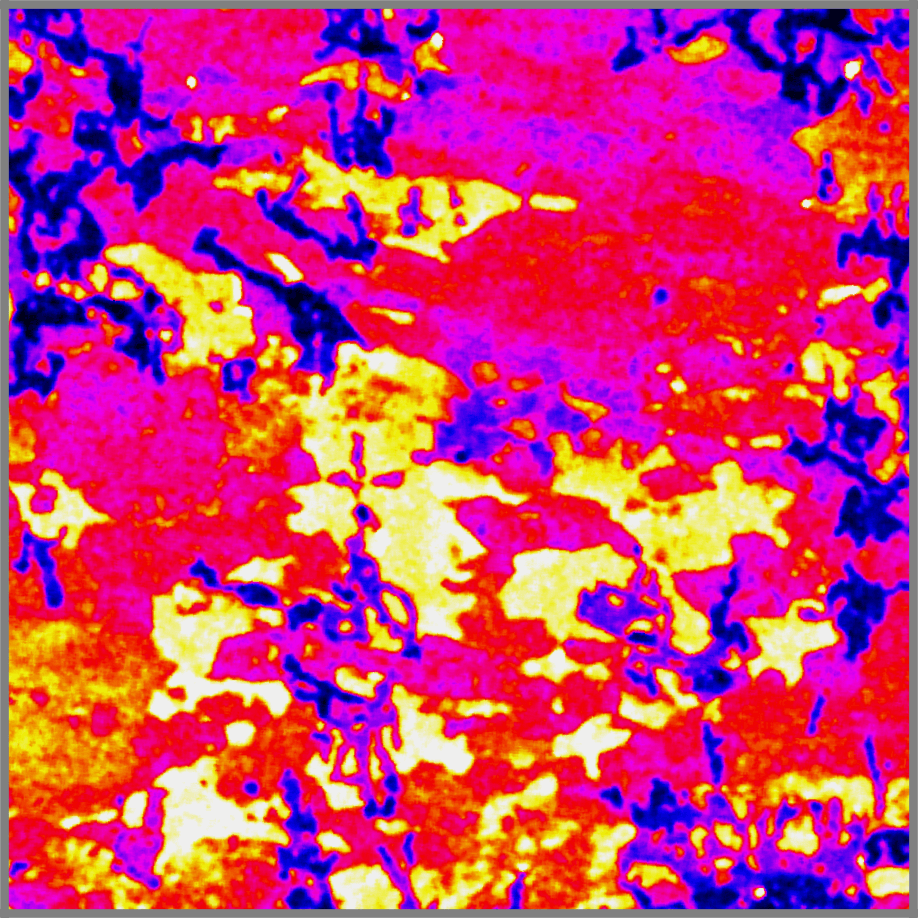};
		\node[anchor=north west,fill opacity=0.75,fill=white] at (rel axis cs:0,1) { {(\textbf{e})} };
		\nextgroupplot[
		axis equal image]
		\addplot graphics[xmin=-25,ymin=-25,xmax=25,ymax=25] {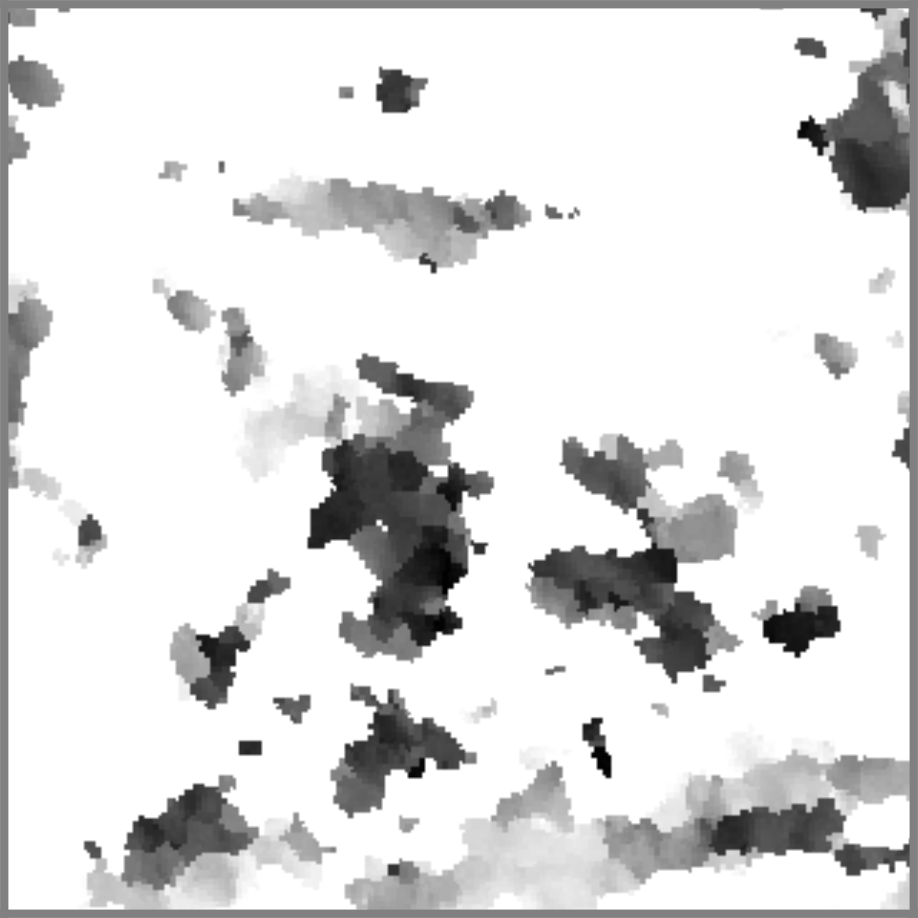};
		\node[anchor=north west,fill opacity=0.75,fill=white] at (rel axis cs:0,1) { {(\textbf{f})} };
		\end{groupplot}
		\node[anchor=south,inner sep=0pt] at (ebsdLegend1) (nodeEbsd1)
		{\includegraphics[width=0.075\textwidth]{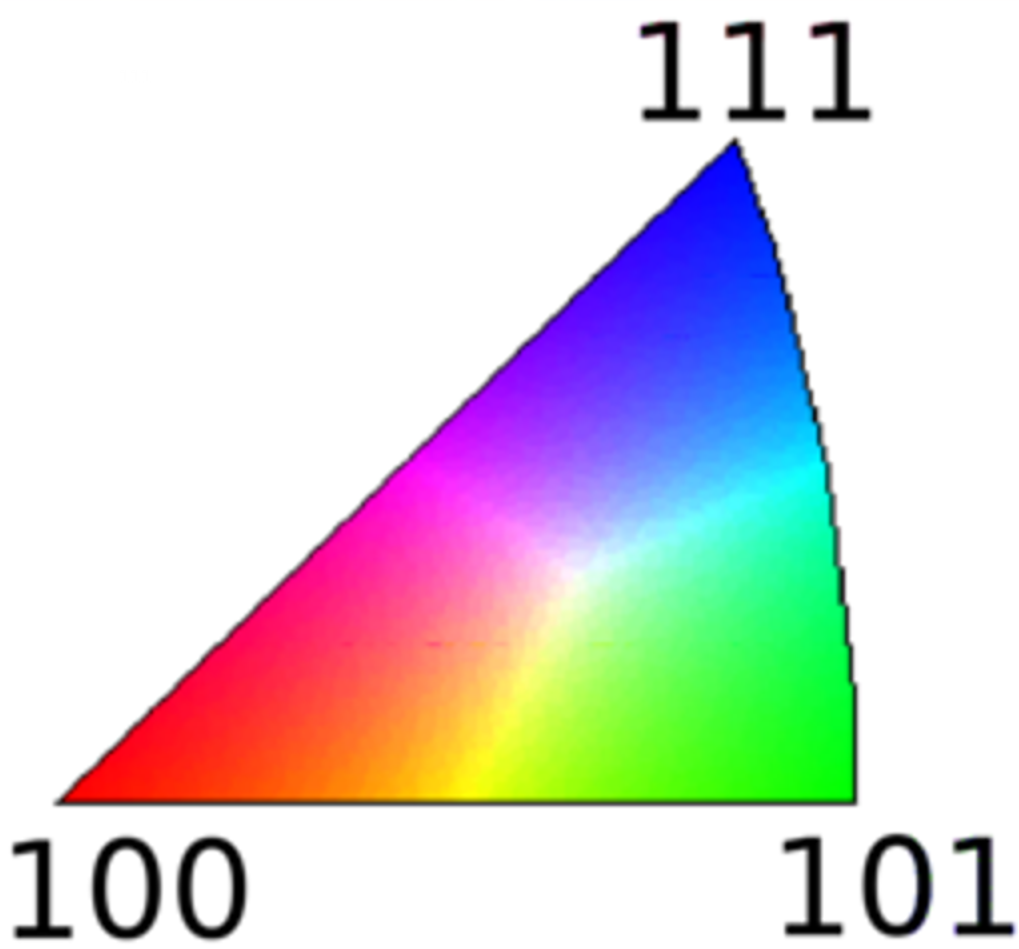}};
		\end{tikzpicture}
	%
	\caption[foo_bar]{Exact overlays of the EBSD IPF$-z$ maps (a, d), sum-filter intensity maps for the Raman peak corresponding to the O-W stretching vibrations (b, e) and the global misorientation angle relative to \hkl{111} substrate orientations (c, f) for: (a, b, c) Sample~\#1~($p_{{\mathrm{Ar-O}}_2}=0.12$~bar) and (d, e, f) Sample~\#2 ($p_{{\mathrm{Ar-O}}_2}=0.22$~bar) respectively (short oxidation at~$T=400$~{\textdegree}C).}
	\label{fig:Raman_EBSD_overlay}
\end{figure*}

The results of applying the sum spectral filter~($770$--$820$~$\mathrm{cm}^{-1}$) to the Raman spectral data for the central region of less oxidised samples~(Fig.~\ref{fig:fiducial}) along with the EBSD grain orientation maps obtained for the same regions prior to oxidation are shown in~Fig.~\ref{fig:Raman_EBSD_overlay}. Oxidation is preferential on tungsten grains with IPF$-z$ orientations close to~\hkl{111} for all samples oxidised at chamber pressures below atmospheric -- as evidenced by the map of misorientation angles relative to the \hkl{111} pole. For grains with a misorientation above approximately~$10^{\circ}$, the Raman peak-filter intensity decreases, indicating a thinner oxide scale, a different oxide phase, or both. The same observation was made when comparing two sets of point spectra acquired directly from the oxide grown on~\hkl{111} and \hkl{001} oriented grains~(Fig.~\ref{fig:Sample3_4_Raman_spectra_texture_comparision}).   
Referring to the recent observation of \citet{Preferential1}, a different trend was observed: tungsten oxide grown on \hkl{111} oriented substrate grains was thinner than on either~\hkl{001} or~\hkl{110} substrate orientations, at~$T=600$~{\textdegree}C for $t=30$~min or at~$T=450$~{\textdegree}C for $t=96$~h. The difference in oxidation conditions should be taken into account; for experiments reported in~\cite{Preferential1} the oxide scale is thicker.

\begin{figure*}[!ht]
	\centering
	\captionsetup{singlelinecheck=off}
	\begin{tikzpicture}
	\begin{groupplot}
	[group style={
		group name=maps1,
		group size=3 by 2,
		xlabels at=edge bottom,
		xticklabels at=edge bottom,
		ylabels at=edge left,
		yticklabels at=edge left,
		vertical sep=10pt,	
		horizontal sep=10pt,	
	},
	xtick align=outside,
	ytick align=outside,
	xtick pos=left,
	ytick pos=left,
	scale only axis,
	xmin=-25,
	xmax=25,
	ymin=-25,
	ymax=25,
	xlabel={$x$~($\mu\rm{m}$)},
	ylabel={$y$~($\mu\rm{m}$)},
	xtick={-20,-10,...,20},
	ytick={-20,-10,...,20},
	axis on top=true,
	width=0.2\textwidth,
	]
	\nextgroupplot[axis equal image]
	\coordinate (ebsdLegend1) at (rel axis cs:0.5,1.03);	
	\addplot graphics[xmin=-25,ymin=-25,xmax=25,ymax=25] {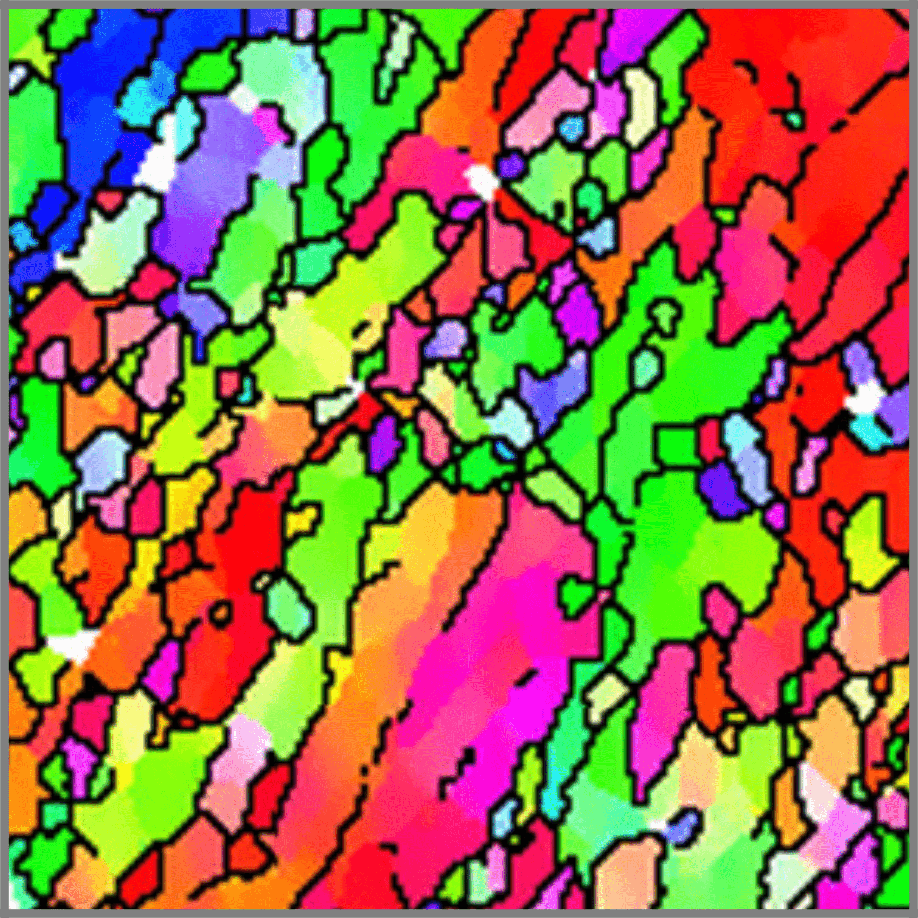};
	\node[anchor=north east,fill opacity=0.75,fill=white] at (rel axis cs:1,1) { {(\textbf{a})} };
	\nextgroupplot[
		axis equal image,
		colorbar horizontal,
		colormap={raman}{color(0cm)=(black) color(1cm)=(blue) color(2cm)=(magenta) color(3cm)=(red) color(4cm)=(yellow) color(5cm)=(white) },
		colorbar style={
			title=Peak-filter intensity,
			at={(0.5,1.2)},anchor=south,
			xticklabel pos=lower,
			xticklabel style={
				/pgf/number format/.cd,
				fixed,
				fixed zerofill,
			},
			xtick={0.1,0.9},
			xticklabels={low,high},
		},
		every colorbar/.append style={height=0.25cm}
		]
	\addplot graphics[xmin=-25,ymin=-25,xmax=25,ymax=25] {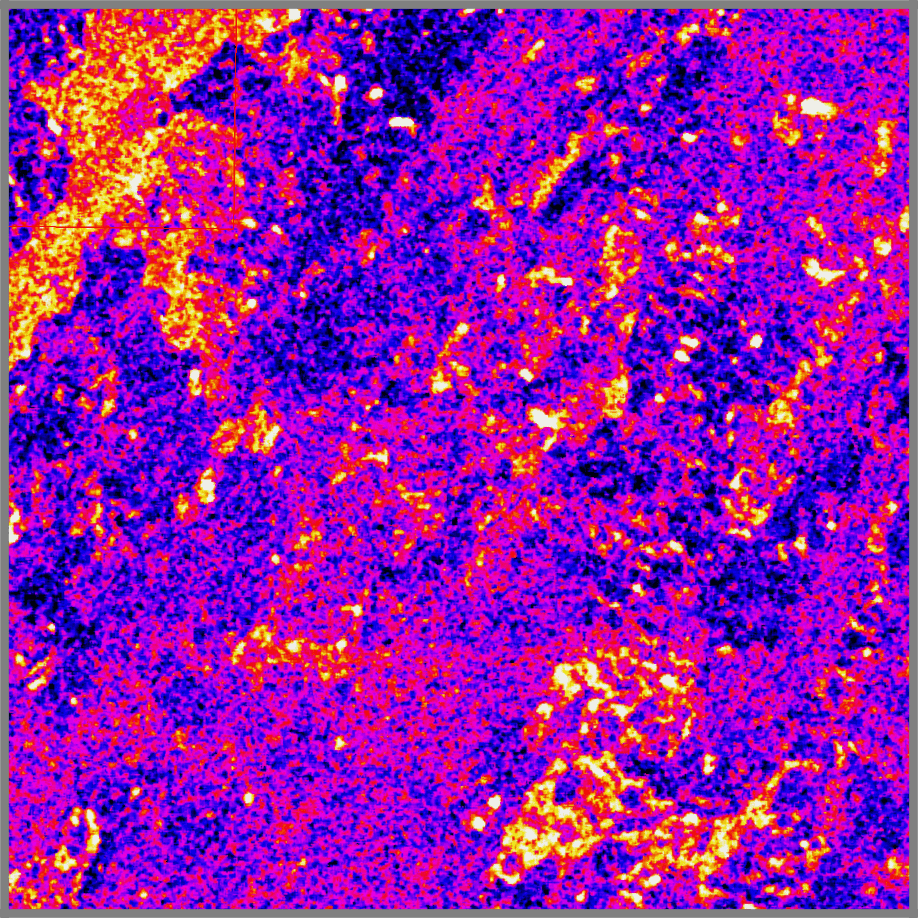};
	\node[anchor=north east,fill opacity=0.75,fill=white] at (rel axis cs:1,1) { {(\textbf{b})} };
	\nextgroupplot[
	axis equal image,
	colorbar horizontal,
	colormap/blackwhite,
	colorbar style={
		title=Misorientation angle,
		at={(0.5,1.2)},anchor=south,
		xticklabel pos=lower,
		point meta min=0.0,
		point meta max=25.0,
		xtick={0.0,5.0,...,25},
	},
	every colorbar/.append style={height=0.25cm}
	]
	\addplot graphics[xmin=-25,ymin=-25,xmax=25,ymax=25] {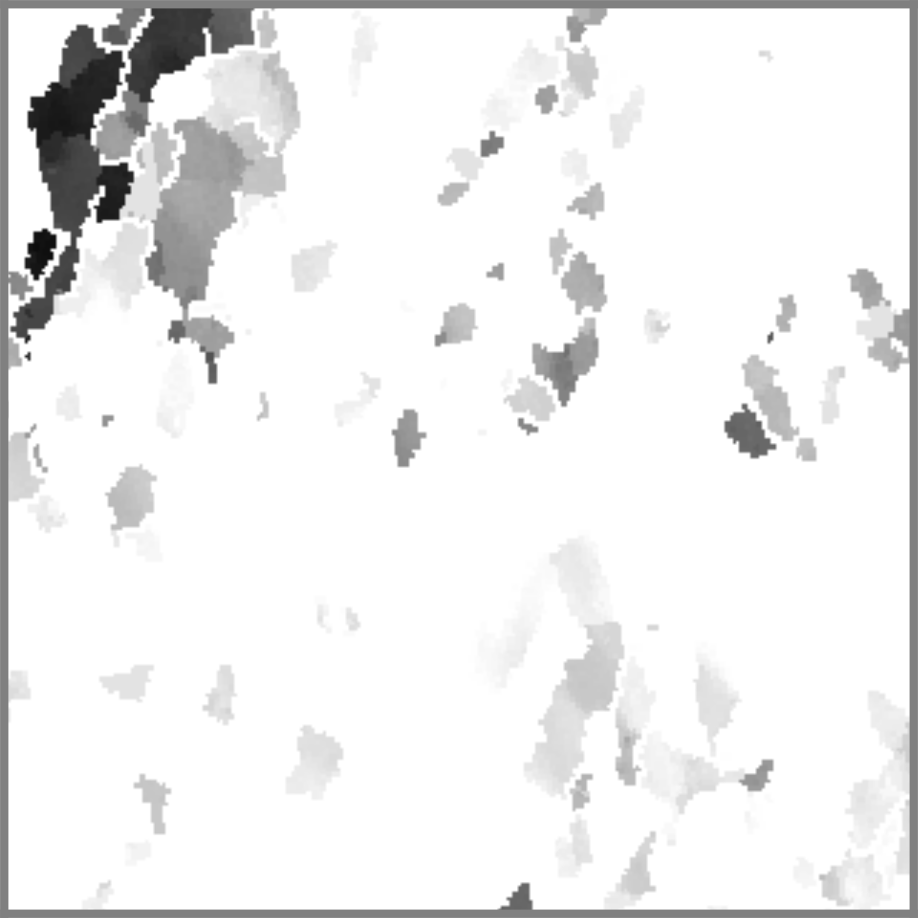};
	\node[anchor=north east,fill opacity=0.75,fill=white] at (rel axis cs:1,1) { {(\textbf{c})} };
	\nextgroupplot[
	axis equal image]
	\coordinate (ebsdLegend2) at (rel axis cs:1,0);	
	\addplot graphics[xmin=-25,ymin=-25,xmax=25,ymax=25] {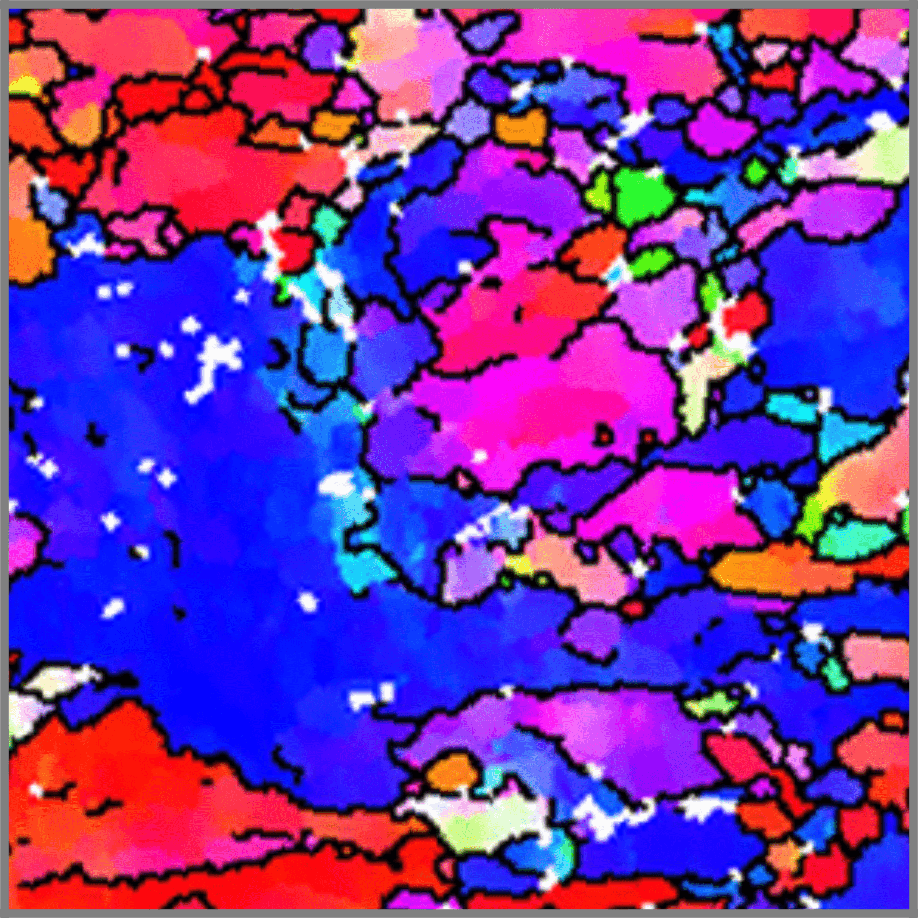};
	\node[anchor=north west,fill opacity=0.75,fill=white] at (rel axis cs:0,1) { {(\textbf{d})} };		
	\nextgroupplot[
	axis equal image]
	\addplot graphics[xmin=-25,ymin=-25,xmax=25,ymax=25] {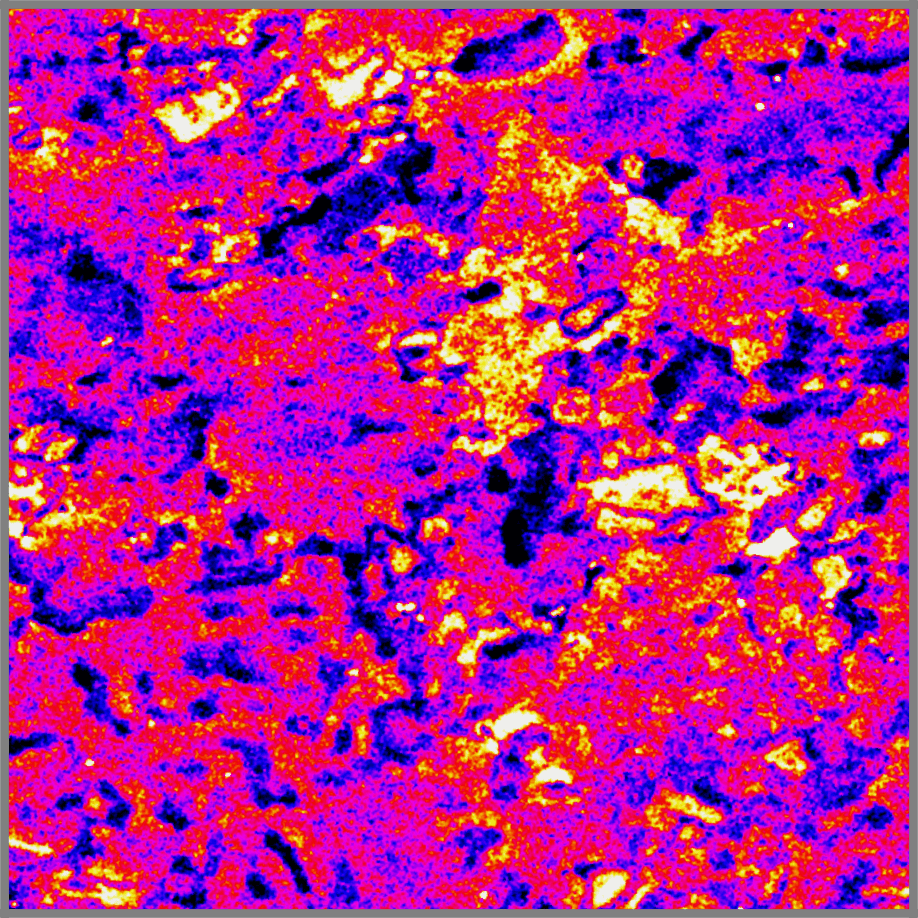};
	\node[anchor=north west,fill opacity=0.75,fill=white] at (rel axis cs:0,1) { {(\textbf{e})} };
	\nextgroupplot[
	axis equal image]
	\addplot graphics[xmin=-25,ymin=-25,xmax=25,ymax=25] {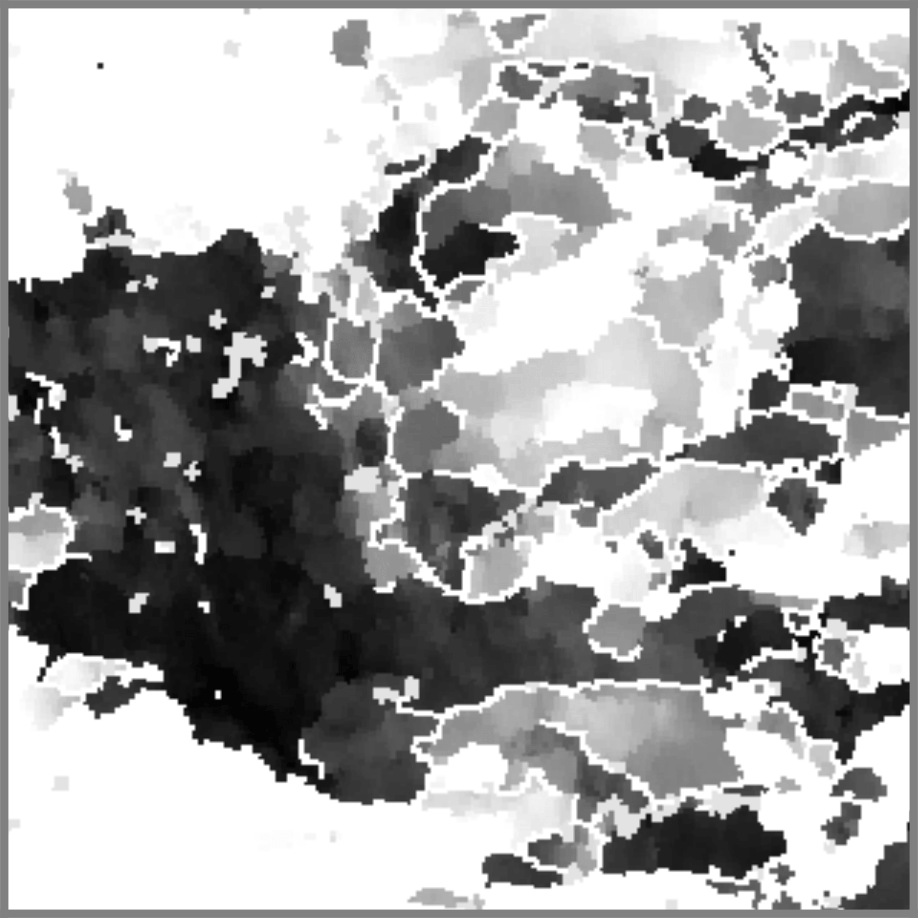};
	\node[anchor=north west,fill opacity=0.75,fill=white] at (rel axis cs:0,1) { {(\textbf{f})} };
	\end{groupplot}
	\node[anchor=south,inner sep=0pt] at (ebsdLegend1) (nodeEbsd1)
	{\includegraphics[width=0.075\textwidth]{Figures/Results/EBSD_bar.png}};
	\end{tikzpicture}
	%
	\caption[foo_bar]{Exact overlays of the EBSD IPF$-z$ maps (a, d), sum-filter intensity maps for the Raman peak corresponding to the O-W stretching vibrations (b, e) and the global misorientation angle relative to \hkl{111} substrate orientations (c, f) for: (a, b, c) Sample~\#3~($t = 20$~min, $p_{{\mathrm{Ar-O}}_2}=1$~bar) and (d, e, f) Sample~\#7 ($t = 72$~h, $p_{{\mathrm{Ar-O}}_2}=1$~bar) respectively (longer oxidation at~$T=400$~{\textdegree}C).}
	\label{fig:Raman_EBSD_overlay_growth}
\end{figure*}

Raman maps gathered from samples oxidised at atmospheric pressure~($p_{\rm{Ar}-\rm{O}_2}=1.0 \ \rm{bar}$) for time intervals between~$t=20$~min to~$t \approx 1$~h also show dominance of the oxide grown on substrate orientations closer to~\hkl{111}~(Fig.~\ref{fig:Raman_EBSD_overlay_growth} (a)--(c)). However, the calculated misorientation range for regions with an enhanced sum-filter intensity is larger~(up to~$25$\textdegree) than for shorter oxidation runs, indicating a weaker preference for the~\hkl{111} substrate orientations compared to short oxidation runs at pressures below atmospheric~(Fig.~\ref{fig:Raman_EBSD_overlay}). In addition, above-average sum-filter intensities were occasionally noticed for substrate orientations closer to the~\hkl{110} pole~(Fig.~\ref{fig:Raman_EBSD_overlay_growth} (a) -- (c)).
For the longest oxidation run~($t=72 \ \rm{h}$) regions of~\hkl{111} type orientations showed below average sum-filter intensity; the highest intensity was observed for a range of substrate orientations between~\hkl{311} and~\hkl{211}~(Fig.~\ref{fig:Raman_EBSD_overlay_growth} (d) -- (f)), indicating a potential transition of the oxidation preference in favour of a different substrate orientation. Following this logic, a shift in oxidation preference to~\hkl{001} substrate orientations may be expected at longer oxidation runs -- as originally proposed by~\citet{Preferential1}.

\subsection{Tungsten oxide phase analysis}

Phase analysis was performed using Raman spectra acquired with the data from the conventional detector~(high intensity, low noise, raster mode) and with the EMCCD detector~(fast acquisition mode). In both cases, averaging was used. In the raster mode, points corresponding to either~\hkl{111} or~\hkl{001} orientations of the tungsten surface grains could be selected. Since EMCCD data is inherently more noisy, it could not be used to accurately determine the fine difference in spectral shapes, hence it was only used for approximate phase analysis of the integral spectra. 

\subsubsection{Analysis of Raman spectra corresponding to~\hkl{001} and \hkl{111} substrate orientations in samples oxidised at~$p_{\mathrm{Ar-O}_{2}}$ $< 1.0 \ \rm{bar}$}
\label{sect:raster_patterns}

\begin{figure*}[!ht]
	\centering
	\captionsetup{singlelinecheck=off}
	\begin{tikzpicture}
	\def\ymin{\pgfkeysvalueof{/pgfplots/ymin}}
	\def\ymax{\pgfkeysvalueof{/pgfplots/ymax}}	
	\begin{groupplot}
	[group style={
		group name=smoothing,
		group size=2 by 1,
		xlabels at=edge bottom,
		ylabels at=edge left,
		xticklabels at=edge bottom,
		yticklabels at=edge left,				
	},
	xlabel={rel. wavenumber~(cm$^{-1}$)},
	ylabel={Intensity (counts)},
	xmin=70,
	xmax=1050,
	ymin=330,
	ymax=460,
	xtick align=outside,
	xtick pos=left,
	ytick pos=left,
	scale only axis,
	width=0.3\textwidth,					
	]
	\nextgroupplot[title={\textbf{(a)}~$\hkl{001}$},legend to name=singleLegend2, legend style={at={(0.5,-0.1)},legend columns=4,fill=none,draw=black,anchor=center,align=center,font=\small,thin}] 
	\coordinate (c11) at (rel axis cs:0,1);
	\addplot[name path=A1,color=blue!25!white,mark=\empty,thin]
	table[x=Raw_X_1,y=Sample3_001_Raman] {Data/Averages/Averages_001.csv};
	\addplot[name path=A2,color=blue!25!white,mark=\empty,thin]				
	table[x=Raw_X_2,y=Sample4_001_Raman] {Data/Averages/Averages_001.csv};
	\addplot[name path=B2,color=black,mark=\empty,densely dashed,thin]
	table[x=Baseline_X_2,y=Baseline_Y_2] {Data/Averages/Averages_001.csv};
	\addplot[color=green!50!black,mark=\empty,very thick]
	table[x=Smoothed_X_1,y=Smoothed_Y_1] {Data/Averages/Averages_001.csv};
	\addplot[color=yellow!85!black,mark=\empty,ultra thick]
	table[x=Smoothed_X_2,y=Smoothed_Y_2] {Data/Averages/Averages_001.csv};
	\addplot[color=black,mark=\empty,densely dashed,thin]
	table[x=Baseline_X_1,y=Baseline_Y_1] {Data/Averages/Averages_001.csv};					
	\legend{,Raw data,Baseline,$p=0.12$~bar,$p=0.22$~bar,,};
	\node[align=center,coordinate,pin = {above left:{$A\approx1080$}}] at (axis cs:810,410) { };
	\node[align=center,coordinate,pin = {above left:{$A\approx440$}}] at (axis cs:810,346) { };
	\draw [densely dotted,thin,gray] (axis cs:810,\ymin) -- (810,\ymax);				
	\node[rotate=90,gray,ultra thick] at (rel axis cs:0.8,0.9) { \small{$810 \ \rm{cm}^{-1}$} };	
	\nextgroupplot[title={\textbf{(b)}~$\hkl{111}$}]
	\coordinate (c21) at (rel axis cs:1,1); 
	\addplot[color=blue!25!white,mark=\empty,thin]				
	table[x=Raw_X_2_1,y=Sample4_111_Raman_1] {Data/Averages/Averages_111.csv};
	\node[align=center,coordinate,pin = {above left:{$A\approx1300$}}] at (axis cs:809,430) { };	
	\addplot[color=blue!25!white,mark=\empty,thin]
	table[x=Raw_X_1_1,y=Sample3_111_Raman_1] {Data/Averages/Averages_111.csv};
	\node[align=center,coordinate,pin = {above left :{$A\approx950$}}] at (axis cs:809,370) { };	
	\addplot[color=yellow!85!black,mark=\empty,ultra thick]
	table[x=Raw_X_2_1,y=Smoothed_Y_2_1] {Data/Averages/Averages_111.csv};	
	\addplot[color=black,mark=\empty,densely dashed, thin]
	table[x=Raw_X_2_1,y=Baseline_Y_2_1] {Data/Averages/Averages_111.csv};
	\addplot[color=green!50!black,mark=\empty,very thick]
	table[x=Raw_X_1_1,y=Smoothed_Y_1_1] {Data/Averages/Averages_111.csv};
	\addplot[color=black,mark=\empty,densely dashed, thin]
	table[x=Raw_X_1_1,y=Baseline_Y_1_1] {Data/Averages/Averages_111.csv};	
	\draw [densely dotted,thin,gray] (axis cs:809,\ymin) -- (809,\ymax);		
	\node[rotate=90,gray,ultra thick] at (rel axis cs:0.8,0.9) { \small{$809 \ \rm{cm}^{-1}$} };			
	\end{groupplot}
	\coordinate (c31) at ($(c11)!.5!(c21)$);
	\node[below] at (c31 |- current bounding box.south)
	{\pgfplotslegendfromname{singleLegend2}};
	\end{tikzpicture}
	\caption[foo_bar]{Full Raman spectra acquired with a~$1800 \ \rm{g}/\rm{cm}^2$ grating, averaged over~$N=25$ regions corresponding to the (a)~\hkl{001} and (b)~\hkl{111} substrate orientations on samples oxidised at chamber pressures below atmospheric. The area of the main peak~$A$~[cm$^{-1}$] is shown for each spectrum.}
	\label{fig:Sample3_4_Raman_spectra_texture_comparision}
\end{figure*}
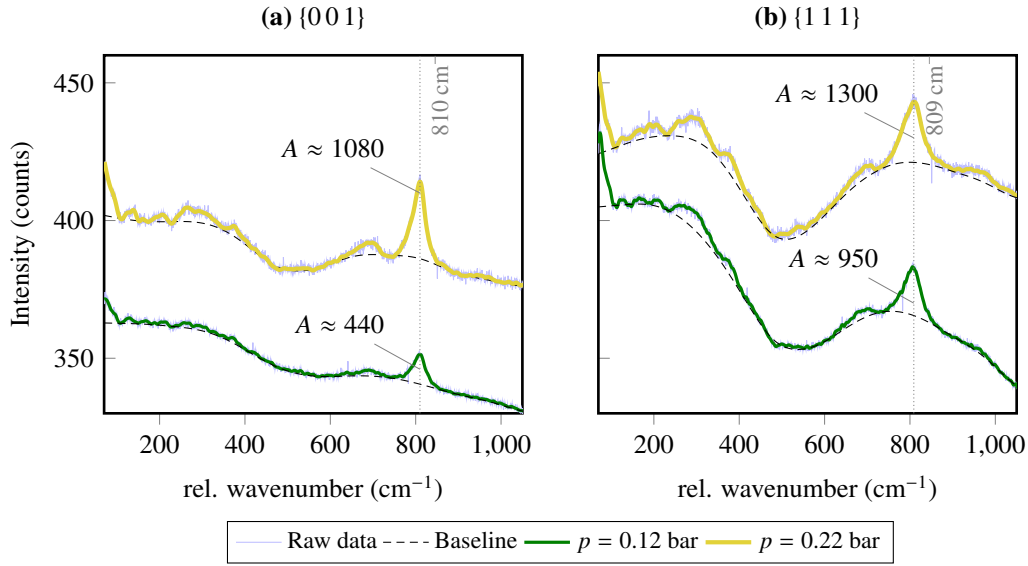

Raman data samples were gathered with a conventional detector from~$N=25$ oxidised regions corresponding to substrate orientations within a~$10${\textdegree} proximity to~\hkl{111} and~\hkl{001} poles. This procedure was repeated for both samples oxidised at below atmospheric pressure~($p_{\mathrm{Ar-O}_{2}} = 0.12$~bar and $p_{\mathrm{Ar-O}_{2}} = 0.22$~bar). The sample-averaged spectra representative to the oxide grown on~\hkl{111} and~\hkl{001} substrate orientations were calculated~(Fig.~\ref{fig:Sample3_4_Raman_spectra_texture_comparision}).
As expected, at higher oxygen partial pressures the area of the main peak~(defined previously by the sum-filter intensity) increases, reflecting the growth of the oxide layer. However, for the oxide grown on~\hkl{111} substrate orientation the main peak area increases only by a factor of~$1.37$, while for the \hkl{001} orientation the growth factor is~$2.45$. This is mainly due to the difference in the full-width half-maximum (FWHM) values; the~$810 \ \rm{cm}^{-1}$ peak corresponding to the~\hkl{001} orientation at~$p_{\mathrm{Ar-O}_{2}} = 0.22 \ \rm{bar}$ is sharper compared to the~$807$--$809 \ \rm{cm}^{-1}$ peak for the~\hkl{111} orientation. The broadening in spectral shape could indicate a phase transformation of the oxide grown on~\hkl{001} substrate orientations. 
 
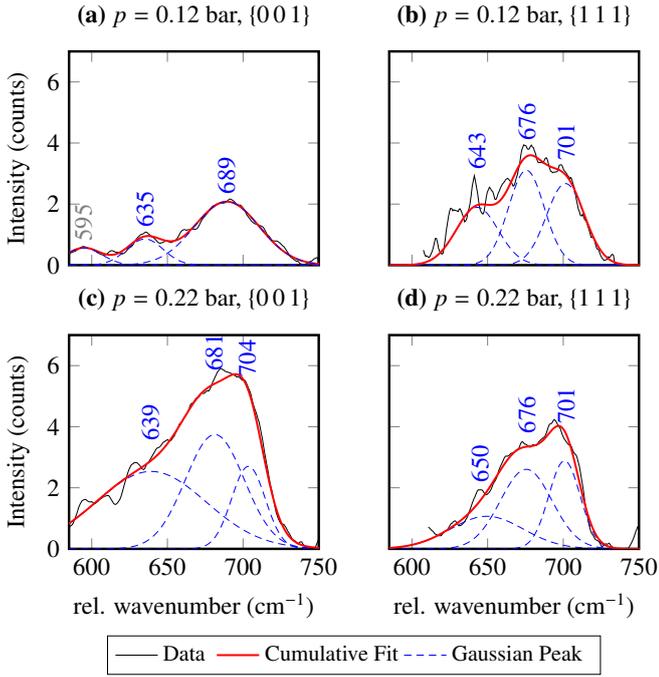
\begin{figure}[!ht]
	\centering
	\captionsetup{singlelinecheck=off}
	\resizebox{1.0\columnwidth}{!}{%
	\begin{tikzpicture}
	\begin{groupplot}
	[group style={
		group name=references,
		group size=2 by 2,
		xlabels at=edge bottom,
		xticklabels at=edge bottom,
		ylabels at=edge left,
		yticklabels at=edge left,
	},
	xlabel={rel. wavenumber~(cm$^{-1}$)},
	ylabel={Intensity (counts)},
	xmin=585,
	xmax=750,
	ymin=0,
	ymax=7,
	xtick distance=50,
	ytick distance=2,
	scale only axis,
	legend style={font=\small},
	width=0.4\columnwidth,
	]
	\nextgroupplot[title={\textbf{(a)}~$p=0.12$~bar, $\hkl{001}$},legend to name=singleLegend1, legend style={at={(0.5,-0.1)},legend columns=4,fill=none,draw=black,anchor=center,align=center,thin}]
	\coordinate (c1) at (rel axis cs:0,1);
	\coordinate (peakPosition1) at (axis cs:635.193811,1.85);
	\coordinate (peakPosition2) at (axis cs:689.27,2.95);
	\coordinate (peakPosition3) at (axis cs:594.75,1.4);
	\addplot[color=black,mark=\empty,thin]
	table[x index=0, y index=1] {Data/Sample3_001.txt};
	\addlegendentry{ Data };
	\node[rotate=90,blue,ultra thick] at (peakPosition1) { {$635$} };
	\node[rotate=90,blue,ultra thick] at (peakPosition2) { {$689$} };		
	\node[rotate=90,gray,ultra thick] at (peakPosition3) { {$595$} };
	\addplot[domain=585:760,samples=200,mark=\empty, red, thick] 
	{ f(x, 16.9705461, 11.9620013, 594.750665) + f(x, 111.895146, 21.4627697, 689.274118) + f(x, 25.7864960, 12.0259574, 635.193811)};				
	\addlegendentry{Cumulative Fit};		
	\addplot[domain=585:760,samples=200,mark=\empty, blue,thin,densely dashed] 
	{ f(x, 25.7864960, 12.0259574, 635.193811) };	
	\addlegendentry{ Gaussian Peak };
	\addplot[domain=585:760,samples=200,mark=\empty, blue,thin,densely dashed] 
	{ f(x, 111.895146, 21.4627697, 689.274118) };			
	\addplot[domain=585:760,samples=200,mark=\empty, blue,thin,densely dashed] 
	{ f(x, 16.9705461, 11.9620013, 594.750665) };	
	\nextgroupplot[title={\textbf{(b)}~$p=0.12$~bar, $\hkl{111}$}]
	\coordinate (c2) at (rel axis cs:1,1); 
	\coordinate (peakPosition4) at (axis cs:643,3.75);
	\coordinate (peakPosition5) at (axis cs:676,4.7);
	\coordinate (peakPosition6) at (axis cs:702,4.05);
	\addplot[color=black,mark=\empty,thin]
	table[x index=0, y index=1] {Data/Sample4_001.txt};	
	\node[rotate=90,blue,ultra thick] at (peakPosition4) { {$643$} };
	\node[rotate=90,blue,ultra thick] at (peakPosition5) { {$676$} };		
	\node[rotate=90,blue,ultra thick] at (peakPosition6) { {$701$} };		
	\addplot[domain=585:760,samples=200,mark=\empty, blue,thin,densely dashed] 
	{ g(x, 92.21699, 27.892, 675.64979) };	
	\addplot[domain=585:760,samples=200,mark=\empty, blue,thin,densely dashed] 
	{ g(x, 66.19936, 32.84562, 643.13716) };			
	\addplot[domain=585:760,samples=200,mark=\empty, blue,thin,densely dashed] 
	{ g(x, 84.22106, 29.61277, 701.45903) };			
	\addplot[domain=585:760,samples=200,mark=\empty, red,thick] 
	{ g(x, 84.22106, 29.61277, 701.45903) + g(x, 66.19936, 32.84562, 643.13716) + g(x, 92.21699, 27.892, 675.64979) };			
	\nextgroupplot[title={\textbf{(c)}~$p=0.22$~bar, $\hkl{001}$}]
	\coordinate (peakPosition4) at (axis cs:639,4.15);
	\coordinate (peakPosition5) at (axis cs:681,6.45);
	\coordinate (peakPosition6) at (axis cs:703,6.25);
	\addplot[color=black,mark=\empty,thin]
	table[x index=0, y index=1] {Data/Sample3_111.txt};
	\node[rotate=90,blue,ultra thick] at (peakPosition4) { {$639$} };
	\node[rotate=90,blue,ultra thick] at (peakPosition5) { {$681$} };		
	\node[rotate=90,blue,ultra thick] at (peakPosition6) { {$704$} };	
	\addplot[domain=585:760,samples=200,mark=\empty, blue,thin,densely dashed] 
	{ g(x, 230.87577, 85.5716, 639.21826) };	
	\addplot[domain=585:760,samples=200,mark=\empty, blue,thin,densely dashed] 
	{ g(x, 182.14783, 45.71567, 681.34225) };			
	\addplot[domain=585:760,samples=200,mark=\empty, blue,thin,densely dashed] 
	{ g(x, 75.5051, 26.25646, 703.53444) };			
	\addplot[domain=585:760,samples=200,mark=\empty, red,thick] 
	{ g(x, 75.5051, 26.25646, 703.53444) + g(x, 182.14783, 45.71567, 681.34225) + g(x, 230.87577, 85.5716, 639.21826) };					
	\nextgroupplot[title={\textbf{(d)}~$p=0.22$~bar, $\hkl{111}$ }]	
	\coordinate (peakPosition7) at (axis cs:645,2.8);
	\coordinate (peakPosition8) at (axis cs:676,4.4);
	\coordinate (peakPosition9) at (axis cs:702,4.7);
	\addplot[color=black,mark=\empty,thin]
	table[x index=0, y index=1] {Data/Sample4_111.txt};
	\node[rotate=90,blue,ultra thick] at (peakPosition7) { {$650$} };
	\node[rotate=90,blue,ultra thick] at (peakPosition8) { {$676$} };		
	\node[rotate=90,blue,ultra thick] at (peakPosition9) { {$701$} };		
	\addplot[domain=585:760,samples=200,mark=\empty, blue,thin,densely dashed] 
	{ g(x, 111.39378, 40.18267, 675.5197) };	
	\addplot[domain=585:760,samples=200,mark=\empty, blue,thin,densely dashed] 
	{ g(x, 74.61519, 24.3142, 700.8187) };			
	\addplot[domain=585:760,samples=200,mark=\empty, blue,thin,densely dashed] 
	{ g(x, 64.5296, 56.87395, 649.51966) };			
	\addplot[domain=585:760,samples=200,mark=\empty, red,thick] 
	{ g(x, 111.39378, 40.18267, 675.5197) + g(x, 74.61519, 24.3142, 700.8187) + g(x, 64.5296, 56.87395, 649.51966) };					
	\end{groupplot}	
	\coordinate (c3) at ($(c1)!.5!(c2)$);
	\node[below] at (c3 |- current bounding box.south)
	{\pgfplotslegendfromname{singleLegend1}};
	\end{tikzpicture}
	}
	\caption[foo_bar]{\label{fig:Texture phase analysis} Peak positions obtained via a deconvolution procedure on the~$585-750 \ \rm{cm}^{-1}$ part of the Raman spectra corresponding to~\hkl{001} and \hkl{111} substrate orientations in: \begin{enumerate*}[label=(\alph*)]
			\item and \item -- Sample~\#1 ($p_{\textrm{Ar-O}_{2}}=0.12$~bar, $t = 20 \ \rm{min}$);
			\item and \item -- Sample~\#2 ($p_{\textrm{Ar-O}_{2}}=0.22$~bar, $t = 20 \ \rm{min}$).
	\end{enumerate*}}
\end{figure}

To verify this, further analysis focused on the~$585$--$750 \ \rm{cm}^{-1}$ part of the Raman spectra -- the characteristic range for determining structural differences -- and the same deconvolution procedure outlined in Section \ref{sect:data_processing} was applied~(Fig.~\ref{fig:Texture phase analysis}).
The oxide developed on~\hkl{001} substrate orientations at~$p_{\textrm{Ar-O}_{2}}=0.12$~bar exhibits Raman peaks at~$595 \ \rm{cm}^{-1}$~(a possible artefact of baseline subtraction), $635 \ \rm{cm}^{-1}$, and $689 \ \rm{cm}^{-1}$. The spectral shape in this wavenumber region closely resembles that of~$h-$WO$_3$~(Fig.~\ref{fig:reference_spectra}), and the peak positions match up with respective reference values~(Table~\ref{tbl:big_reference_table}) within an acceptable error margin. For the~\hkl{111} substrate orientation, the spectral shape is significantly different, with peaks at $639~\mathrm{cm^{-1}}$, $681~\mathrm{cm^{-1}}$, and $704~\mathrm{cm^{-1}}$; these match up well with the reference~$o-$WO$_3$ peak positions~(Table~\ref{tbl:big_reference_table}). Oxidation at~$p_{\textrm{Ar-O}_{2}}=0.22$~bar results in a higher intensity of the Raman signal in the~$585$--$750 \ \rm{cm}^{-1}$ wavenumber region~(bending vibrations) for the oxide on~\hkl{001} substrate orientations. However, the sum-filter intensity for the \hkl{001} type Raman spectrum the~$770$--$820 \ \rm{cm}^{-1}$ wavenumber region (stretching vibrations) is lower compared to that of~\hkl{111} type spectrum. This supports the observations previously made in~Sect.~\ref{sect:maps}: the oxide should be thicker for~\hkl{111} orientations. The spectrum in this region~(Fig.~\ref{fig:Texture phase analysis} (c)) resembles a superposition of $h-$WO$_3$ (Fig.~\ref{fig:Texture phase analysis} (a)) and~$o-$WO$_3$~(Fig.~\ref{fig:Texture phase analysis} (b)) type spectra, possibly indicating stacking of two oxide phases. Raman spectra for the oxides grown on \hkl{111} substrate orientations~(Fig.~\ref{fig:Texture phase analysis} (b), (d)) for both oxidation conditions show roughly the same peaks at~$(646 \pm 4)~\mathrm{cm^{-1}}$, $(676 \pm 1)~\mathrm{cm^{-1}}$, and $(701 \pm 1)~\mathrm{cm^{-1}}$, all characteristic to $o$--WO$_3$.

The formation of metastable $h$--WO$_3$ on \hkl{001} type grains might explain the sluggish growth during the onset of oxidation. The Gibbs free energy cost of its formation relative to the~$o$--WO$_3$ phase thermodynamically stable at~$T=400${\textdegree}C~\cite{PolymorpW} scales with the thickness of the oxide and acts to decrease the stability of the~$h$--WO$_3$ during growth. It is possible that the~$h$--WO$_3$ phase forms preferentially on tungsten grains with~\hkl{001} orientation because of a lower crystal structure mismatch~(as in case of epitaxial growth), thereby reducing the effective interfacial energy of the $h$--WO$_3$ bulk relative to $o$--WO$_3$ bulk interface. However, this negative interfacial energy is independent of the oxide thickness and therefore, the initial reduction in free energy costs is outweighed as the oxide grows thicker. This might explain why the $o$--WO$_3$ phase becomes dominant.

\subsubsection{Phase analysis of the oxide layer formed at atmospheric pressure after~$t > 20 \ \rm{min}$}

The kinetics of the oxide layer growth on tungsten at~$t \gtrapprox 72$~h show a shift of oxidation preference towards~\hkl{001} substrate orientations~(Sect.~\ref{sect:maps}), which cannot be explained by the early~$h-$WO$_3$ to~$o-$WO$_3$ transition. In this section, the long-term phase evolution of the oxide layer is explored.

The oxide film, initially appearing as pale-yellow, starts developing a blueish hue at~$t \gtrapprox 1$~h of oxidation~(illustrated in Fig.~\ref{fig:Topography_histogram_plot} and~Fig.~\ref{fig:Raman_Spectra_Long_Oxidation}). A tungsten blue oxide~(TBO, \cite{TBO}) is known to represent oxygen-deficient tungsten trioxide formed by W atoms in different oxidation states~\cite{weil2013beautiful}. 
\begin{figure}[!ht]
	\centering
	\captionsetup{singlelinecheck=off}
	\captionsetup[subfigure]{justification=centering}
	\begin{subfigure}[t]{0.75\linewidth}
		\centering
	\subcaption{}
		\includegraphics[width=1.0\linewidth]{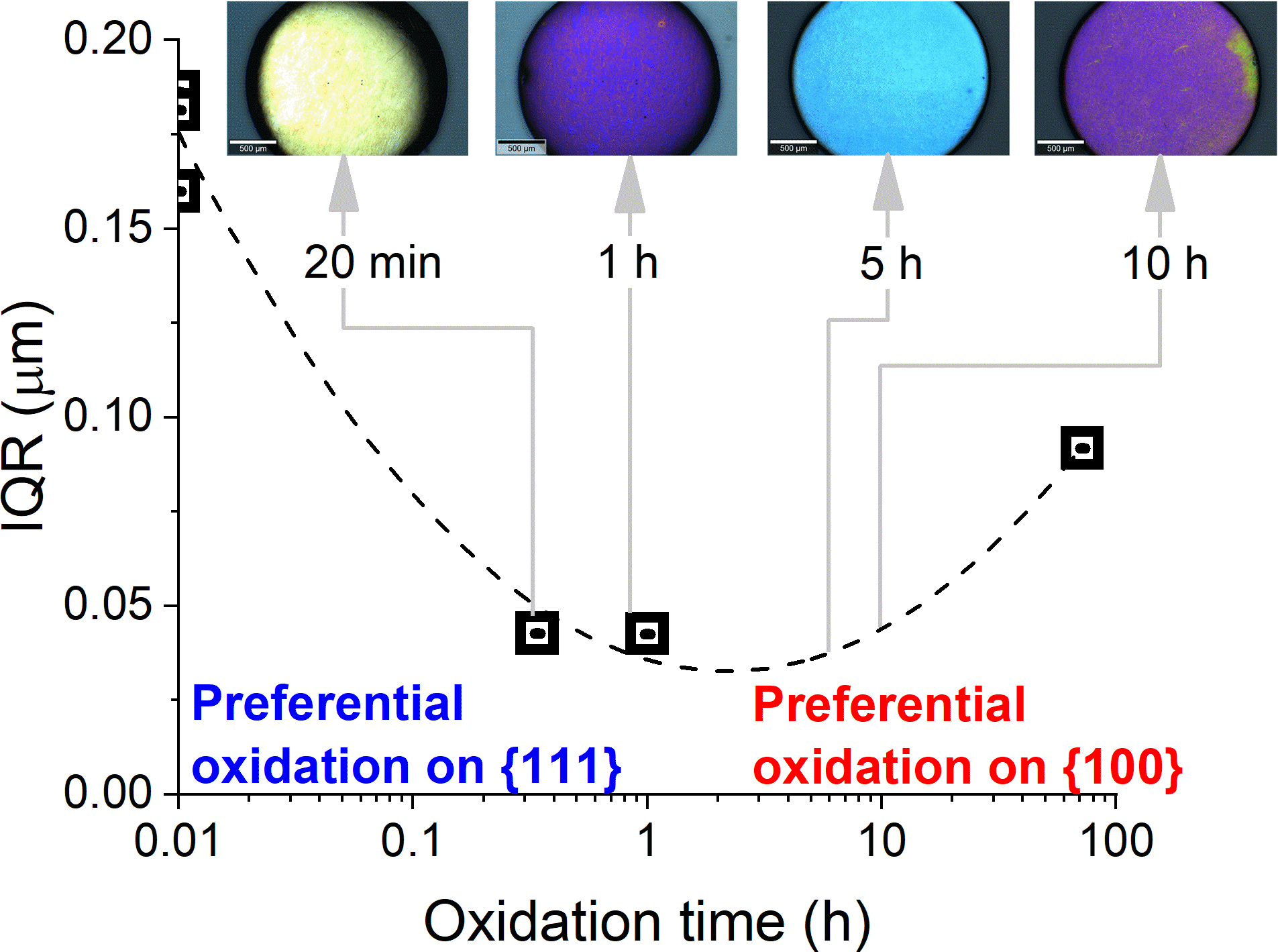}
		\label{fig:Increase_decrease_topography}
	\end{subfigure}
	~
	\begin{subfigure}[t]{0.75\linewidth}
	\centering
	\subcaption{}
	\includegraphics[width=1.0\linewidth]{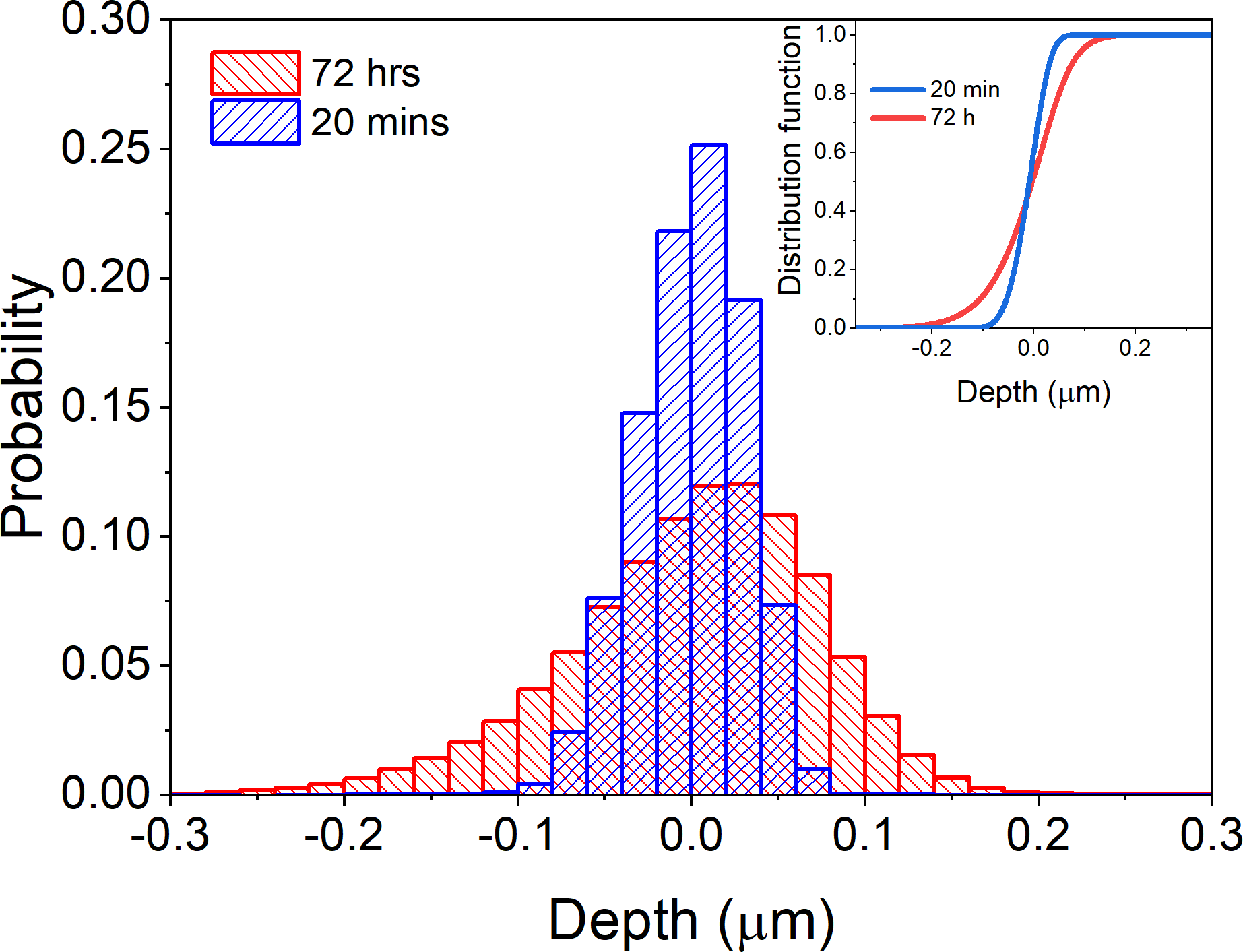}
	\label{fig:Topography_histogram}
	\end{subfigure}%
	\caption[foo_bar]{Results of CLSM image processing showing:  \begin{enumerate*}[label=(\alph*)]
			\item evolution of interquartile range~(IQR) of surface roughness from as-polished tungsten, through oxide initiation to growth, showing an initial drop in topography and subsequent increase. It is inferred that the minimum of~IQR corresponds to reversal of oxidation preference from~\hkl{111} to \hkl{001} tungsten planes;
			\item the distribution of relative surface height in tungsten samples oxidised for $20$~min and $72$~h.
	\end{enumerate*}}
	\label{fig:Topography_histogram_plot}
\end{figure}
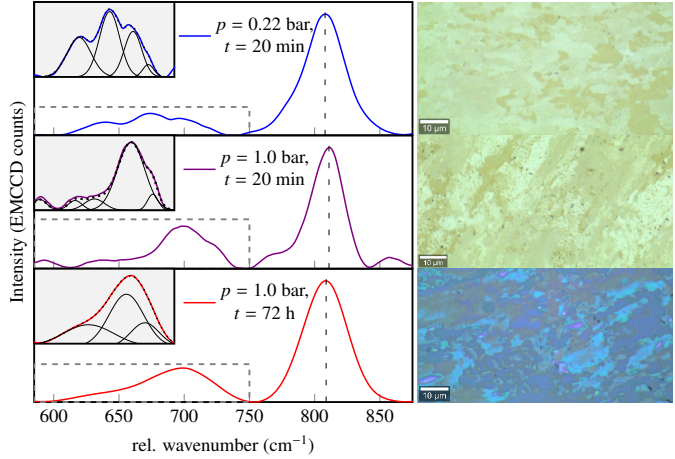
\begin{figure}
	\captionsetup{singlelinecheck=off}
	\centering
	\resizebox{1.0\columnwidth}{!}{%
	\begin{tikzpicture}
		\def\xmin{\pgfkeysvalueof{/pgfplots/xmin}}
		\def\xmax{\pgfkeysvalueof{/pgfplots/xmax}}
		\def\ymin{\pgfkeysvalueof{/pgfplots/ymin}}
		\def\ymax{\pgfkeysvalueof{/pgfplots/ymax}}	
		\begin{groupplot}
		[group style={
			group size=1 by 3,
			xlabels at=edge bottom,
			xticklabels at=edge bottom,
			ylabels at=edge left,
			vertical sep=0pt,
		},
		xlabel={rel. wavenumber (cm$^{-1}$)},
		ylabel={Intensity (counts)},
		xmin=585,
		xmax=875,
		ymin=0,
		xtick pos=left,
		yticklabels={,,},		
		width=0.8\columnwidth,
		height=0.105\textheight,
		scale only axis,	
		ytick=\empty,
		]
		\nextgroupplot[ylabel={},
		legend style={at={(0.565,0.95)},legend columns=1,fill=none,draw=none,anchor=north,align=center}
		]
		\addplot[color=blue,mark=\empty,thick]
		table[x expr=\thisrowno{0}, y expr=\thisrowno{1}] {Data/Full_Spectra.dat};
		\coordinate (insetPosition1) at (rel axis cs:0,1);
		\coordinate (imagePosition1) at (rel axis cs:1.35,0.5);
		\draw [gray, dashed] (axis cs:585,0.0) rectangle (axis cs:750,0.3);	
		\draw [loosely dashed,thin] (axis cs:808,\ymin) -- (808,\ymax);			
		\addlegendentry{ $p=0.22$~bar,\\$t=20$~min }	
		\nextgroupplot[ylabel={Intensity~(EMCCD counts)},
		legend style={at={(0.565,0.95)},legend columns=1,fill=none,draw=none,anchor=north,align=center}
		]
		\coordinate (insetPosition2) at (rel axis cs:0,1);
		\coordinate (imagePosition2) at (rel axis cs:1.35,0.5);		
		\addplot[color=blue!50!red,mark=\empty,thick]
		table[x expr=\thisrowno{2}, y expr=\thisrowno{3}] {Data/Full_Spectra.dat};	
		\addlegendentry{ $p=1.0$~bar,\\$t=20$~min }	
		\draw [gray, dashed] (axis cs:585,0.0) rectangle (axis cs:750,3.25);
		\draw [loosely dashed,thin] (axis cs:811,\ymin) -- (811,\ymax);		
		\nextgroupplot[ylabel={},
		legend style={at={(0.565,0.95)},legend columns=1,fill=none,draw=none,anchor=north,align=center}
		]
		\coordinate (insetPosition3) at (rel axis cs:0,1);
		\coordinate (imagePosition3) at (rel axis cs:1.35,0.5);		
		\addplot[color=red,mark=\empty,thick]
		table[x expr=\thisrowno{6}, y expr=\thisrowno{7}] {Data/Full_Spectra.dat};
		\addlegendentry{ $p=1.0$~bar,\\$t=72$~h }	
		\draw [gray, dashed] (axis cs:585,0.0) rectangle (axis cs:750,0.65);
		\draw [loosely dashed,thin] (axis cs:808.8,\ymin) -- (808.8,\ymax);							
		\end{groupplot}
		\begin{axis}[
		at={(insetPosition1)},anchor={north west},footnotesize,width=0.475\columnwidth,height=0.125\textheight,xmin=585,xmax=750,ymin=0,ytick=\empty,xtick=\empty,axis background/.style={fill=gray!10}]		
		\addplot[color=blue,mark=\empty,thick]
		table[x expr=\thisrowno{0}, y expr=\thisrowno{1}] {Data/Full_Spectra.dat};
		\addplot[domain=585:760,samples=200,mark=\empty, black,very thin] 
		{ g(x, 4.77925, 33.03443, 637.89079) };	
		\addplot[domain=585:760,samples=200,mark=\empty, black,very thin] 
		{ g(x, 6.65958, 27.83028, 673.76318) };	
		\addplot[domain=585:760,samples=200,mark=\empty, black,very thin] 
		{ g(x, 4.02427, 24.15687, 701.33535) };			
		\addplot[domain=585:760,samples=200,mark=\empty, black,very thin] 
		{ g(x, 0.69401, 15.28326, 719.33729) };					
		\addplot[domain=585:760,samples=200,mark=\empty, black,thick, dotted] 
		{ g(x, 0.69401, 15.28326, 719.33729) + g(x, 4.02427, 24.15687, 701.33535) + g(x, 6.65958, 27.83028, 673.76318) + g(x, 4.77925, 33.03443, 637.89079) };	
		\end{axis}
		\begin{axis}[
		at={(insetPosition2)},anchor={north west},footnotesize,width=0.475\columnwidth,height=0.125\textheight,xmin=585,xmax=750,ymin=0,ytick=\empty,xtick=\empty,axis background/.style={fill=gray!10}]		
		\addplot[color=blue!50!red,mark=\empty,thick]
		table[x expr=\thisrowno{2}, y expr=\thisrowno{3}] {Data/Full_Spectra.dat};
		\addplot[domain=585:760,samples=200,mark=\empty, black,very thin] 
		{ g(x, 112.18909, 37.54701, 699.05659) };	
		\addplot[domain=585:760,samples=200,mark=\empty, black,very thin] 
		{ g(x, 10.43052, 14.76462, 724.66712) };	
		\addplot[domain=585:760,samples=200,mark=\empty, black,very thin] 
		{ g(x, 8.26867, 19.24671, 633.16014) };			
		\addplot[domain=585:760,samples=200,mark=\empty, black,very thin] 
		{ g(x, 13.96251, 28.0273, 656.00561) };					
		\addplot[domain=585:760,samples=200,mark=\empty, black, very thin] 
		{ g(x, 9.20108, 19.3478, 591.67191) };	
		\addplot[domain=585:760,samples=200,mark=\empty, black, very thick, dotted] 
		{ g(x, 112.18909, 37.54701, 699.05659) + g(x, 10.43052, 14.76462, 724.66712) + g(x, 8.26867, 19.24671, 633.16014) + g(x, 9.20108, 19.3478, 591.67191) + g(x, 13.96251, 28.0273, 656.00561) };			
		\end{axis}
		\begin{axis}[
		at={(insetPosition3)},anchor={north west},footnotesize,width=0.475\columnwidth,height=0.125\textheight,xmin=585,xmax=750,ymin=0,ytick=\empty,xtick=\empty,axis background/.style={fill=gray!10}]		
		\addplot[color=red,mark=\empty,thick]
		table[x expr=\thisrowno{6}, y expr=\thisrowno{7}] {Data/Full_Spectra.dat};
		\addplot[domain=585:760,samples=200,mark=\empty, black,very thin] 
		{ -0.02668 + g(x, 14.05571, 70.01991, 648.93812) };	
		\addplot[domain=585:760,samples=200,mark=\empty, black,very thin] 
		{ -0.02668 + g(x, 23.03244, 48.12281, 693.53091) };	
		\addplot[domain=585:760,samples=200,mark=\empty, black,very thin] 
		{ -0.02668 + g(x, 8.54522, 39.09776, 715.40475) };			
		\addplot[domain=585:760,samples=200,mark=\empty, black,thick, dotted] 
		{ -0.02668 + g(x, 8.54522, 39.09776, 715.40475) + g(x, 23.03244, 48.12281, 693.53091) + g(x, 14.05571, 70.01991, 648.93812) };	
		\end{axis}		
		\node[inner sep = 0pt] at (imagePosition1) (sample4) {\includegraphics[height=0.105\textheight]{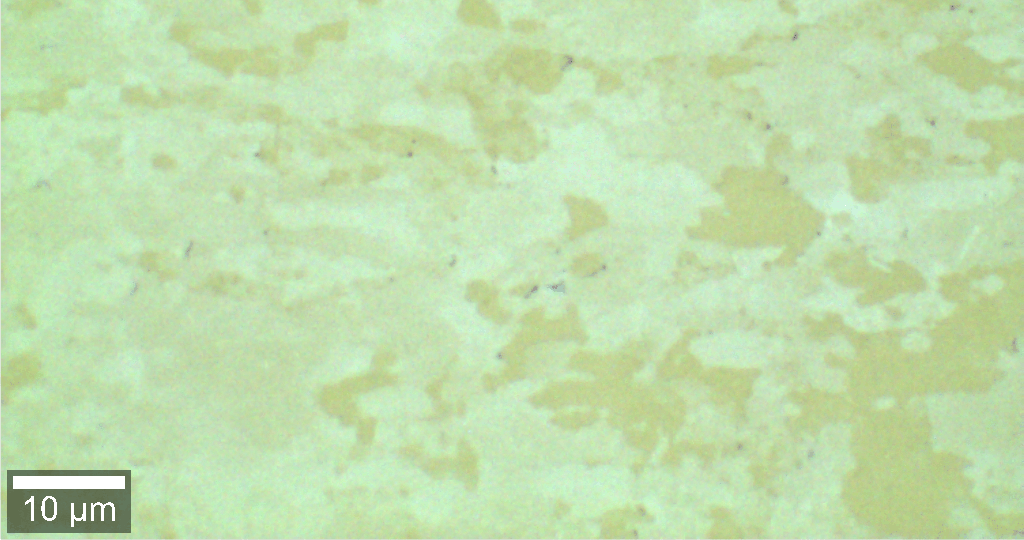}};
		\node[inner sep = 0pt] at (imagePosition2) (20min) {\includegraphics[height=0.105\textheight]{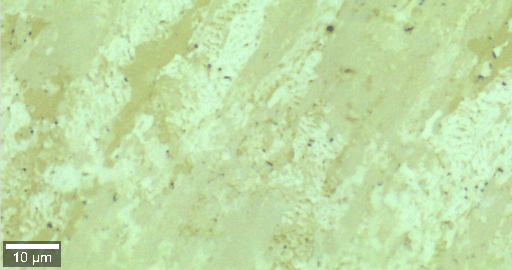}};		
		\node[inner sep = 0pt] at (imagePosition3) (72h) {\includegraphics[height=0.105\textheight]{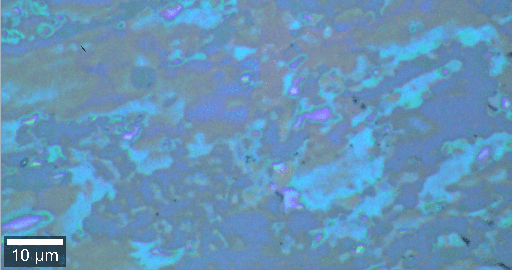}};				
		\end{tikzpicture}
	}	
	\caption[foo_bar]{EMCCD Raman spectra (left column) averaged over large areas displayed by the optical images (right column), showing how the main stretching and bending Raman bands evolve at different stages of oxidation.} 
	\label{fig:Raman_Spectra_Long_Oxidation}
\end{figure}
The oxide layers observed on samples oxidised at~$t > 20$~min are polychromatic, sometimes combining dark-yellow and blue~($t = 1$~h) or light and dark blue, pink, and yellow~($t=72$~h). The colours supposedly correlate with the oxygen concentration~\cite{weil2013beautiful}. The shift in preferential oxidation observed simultaneously with the colour change is likely driven by this change of composition via a diffusion-assisted process. This heterogeneity of colour and oxygen composition in the tungsten oxide layer across the entire sample means that the resulting Raman spectra must be smoothed over contributions from different oxide phases, the Raman activities of which are also different~(Figure \ref{fig:Raman_Spectra_Long_Oxidation}). Therefore, it is impossible to resolve Raman signal from differently-coloured regions, but the principal component of a large Raman EMCCD dataset may be used to give some integral information on the dominant phases. This was carried out with the WiTec TrueComponent Analysis tool on the same dataset reported in Sect.~\ref{sect:maps}. The principal components and the deconvoluted Raman peaks corresponding to the O--W bending vibrations are presented in~Fig.~\ref{fig:Raman_Spectra_Long_Oxidation}. Results of peak deconvolution with a comparison to the references are given in Table~\ref{tbl:Raman_Long_Oxidation}. 

\begin{table}[!ht]
	\centering
	\caption{Deconvolution results for the overlapping peaks corresponding to bending vibrations in Raman spectra~(Fig.~\ref{fig:Raman_Spectra_Long_Oxidation}) of tungsten samples oxidised at~$T = 400 \ $°$ \rm{C}$. The fitting results are compared to the reference peak positions listed in Table~\ref{tbl:big_reference_table}. Where fitting errors~$\Delta > 1 \ \rm{cm}^{-1}$, their values are explicitly shown.}
	\label{tbl:Raman_Long_Oxidation}
	\resizebox{\columnwidth}{!}{%
	\begin{tabular}{lllc}
		\hline
		\multicolumn{1}{c}{Oxidation conditions}     & \multicolumn{2}{c}{\begin{tabular}[c]{@{}c@{}}Peak position \\(cm$^{-1}$)\end{tabular}} &  \multicolumn{1}{c}{\begin{tabular}[c]{@{}c@{}}Phase\end{tabular}}    \\ 
		& Measured & Reference & \\
		\hline
		\multirow{4}{*}{\begin{tabular}[l]{@{}l@{}} $p_{\textrm{Ar}-\textrm{O}_2} = 0.22$~bar \\ $t = 20$~min \end{tabular}} & 637                                                                                          & 640                                                                                     & \multicolumn{1}{c}{$o -$WO$_3$}                                                           \\
		& 674                                                                                          & 680                                                                                     & \multicolumn{1}{c}{$o -$WO$_3$}                                                           \\
		& 701                                                                                          & 704                                                                                     & \multicolumn{1}{c}{$o -$WO$_3$}                                                           \\
		& $719 \pm 1$                                                                                          & 718                                                                                     & \multicolumn{1}{c}{$m -$WO$_3$}                                                           \\ \hline
		\multirow{5}{*}{ \begin{tabular}[l]{@{}l@{}} $p_{\textrm{Ar}-\textrm{O}_2} = 1.0$~bar \\ $t = 20$~min \end{tabular}} & 591                                                                                          & 595                                                                                     & \multicolumn{1}{c}{$m -$WO$_2$}                                                           \\
		& $633 \pm 1$                                                                                          & 640                                      & $o -$WO$_3$                        \\
		& $656 \pm 1$                                                                                          & 660                                                                                     & $m -$WO$_2$                                                                               \\
		& 699                                                                                          & 696                                                                                     & $m -$WO$_2$                                                                               \\
		& 725                                                                                          & 730                                                             & $o -$WO$_3$                                                                                                                                                                                                                 \\ \hline
		\multirow{3}{*}{\begin{tabular}[l]{@{}l@{}} $p_{\textrm{Ar}-\textrm{O}_2} = 1.0$~bar \\ $t = 72$~h \end{tabular}}   & $649 \pm 3$                                                                                          & 646; 653                                                                                & \multicolumn{1}{c}{$h -$WO$_3$}                                                           \\
		& $694 \pm 6$                                                                                          & 690                                                                 & $h -$WO$_3$                                                                               \\
		& $715 \pm 3$                                                                                          & 718                                                                 & {$m-$WO$_3$}                                                                               \\
		\hline
	\end{tabular}%
	}
\end{table}
The averaged spectra calculated from the EMCCD datasets for~$t=20$~min oxidation runs at~$p_{\rm{Ar}-\rm{O}_2}=0.22 \ \rm{bar}$ and~$p_{\rm{Ar}-\rm{O}_2}=1.0 \ \rm{bar}$ respectively were significantly different, although little colour change was observed~(Fig.~\ref{fig:Raman_Spectra_Long_Oxidation}). The results of peak deconvolution~(Table~\ref{tbl:Raman_Long_Oxidation}) show a dominant~$o-$WO$_3$ phase at~$p_{\rm{Ar}-\rm{O}_2}=0.22 \ \rm{bar}$; an additional small peak at~$719$~cm$^{-1}$ could be attributed to the residual native $m-$WO$_3$ oxide stable at room temperature. Increasing oxidation pressure from~$p_{\rm{Ar}-\rm{O}_2}=0.22 \ \rm{bar}$ to $p_{\rm{Ar}-\rm{O}_2}=1.0 \ \rm{bar}$ during the first~$t=20$~min of oxidation leads to a shift in local maximum from~$674 \ \rm{cm}^{-1}$ to~$699 \ \rm{cm}^{-1}$. Deconvolution shows a two-phase composition: a mix of~$o-$WO$_3$ and~$m-$WO$_2$. The results of optical microscopy are inconclusive and more analysis is needed with electron microscopy. 

The phase diagram of the W--O system~\cite{habainy2018} shows that a dual-phase region exists at~$T=400$~{\textdegree}C consisting of a mixture of WO$_3$ and WO$_2$.  However, the phase diagram considers the thermodynamic stability of bulk phases and is not strictly applicable for thin oxide layers in this work; despite this, general comparisons are drawn. As the oxide scale grows, it is expected that while the highest oxide~(WO$_3$) remains at the surface, the deeper oxide layers will develop a gradient of oxygen concentration~\cite{habainy2018,TungstenOxidation2}. The slight colour shift towards brown for some regions of the optical image is then likely attributed to the development of the~$m-$WO$_2$~\cite{weil2013beautiful} phase. At $p_{\rm{Ar}-\rm{O}_2}=1.0 \ \rm{bar}$ and~$T=400$~{\textdegree}C the formation of TBO~(e.g. at~$t \gtrapprox 1$~h) on tungsten metal is unexpected: according to~\citet{WOSystem1}, at~$330 \ \rm{{\degree}C} < T < 484 \ \rm{{\degree}C}$ only the WO$_2$ and~$o-$WO$_3$ are thermodynamically stable, while TBO usually forms by decomposition of tungstic acid~(HWO$_4$)~\cite{TBO}. The Raman spectrum of the sample oxidised at~$t=72$~h shows a well-developed~$808$~cm$^{-1}$ peak characteristic to the~$m-$WO$_3$ phase and additional peaks, which are somewhat close to those found for the~$h-$WO$_3$ phase. Roughly the same spectral shape is obtained for~$t = 1$~h oxidation, where the oxide layer~(appearing as partially violet) should be much thinner than for~$t=72$~h oxidation, but the spectral shape is broader, indicating a mixture of an array of different oxides. The presence of~$m-$WO$_2$ is no longer observed and this can be attributed to the restricted penetration depth of the laser, which is unable to probe the lower oxides located closer to the substrate material. Hence, additional experiments are required to establish the stacked nature of these phases, which is believed to be responsible for the reversal of oxidation kinetics. These stacked oxides might contain the TBO phase, acting as an interface between~$m-$WO$_2$ and ~$o-$WO$_3$, mediating the oxidation rate via diffusion-assisted processes.
	
\subsection{Correlating EBSD and confocal topography maps}

\begin{figure}
	\centering
	\captionsetup{singlelinecheck=off}
	\includegraphics[width=0.75\linewidth]{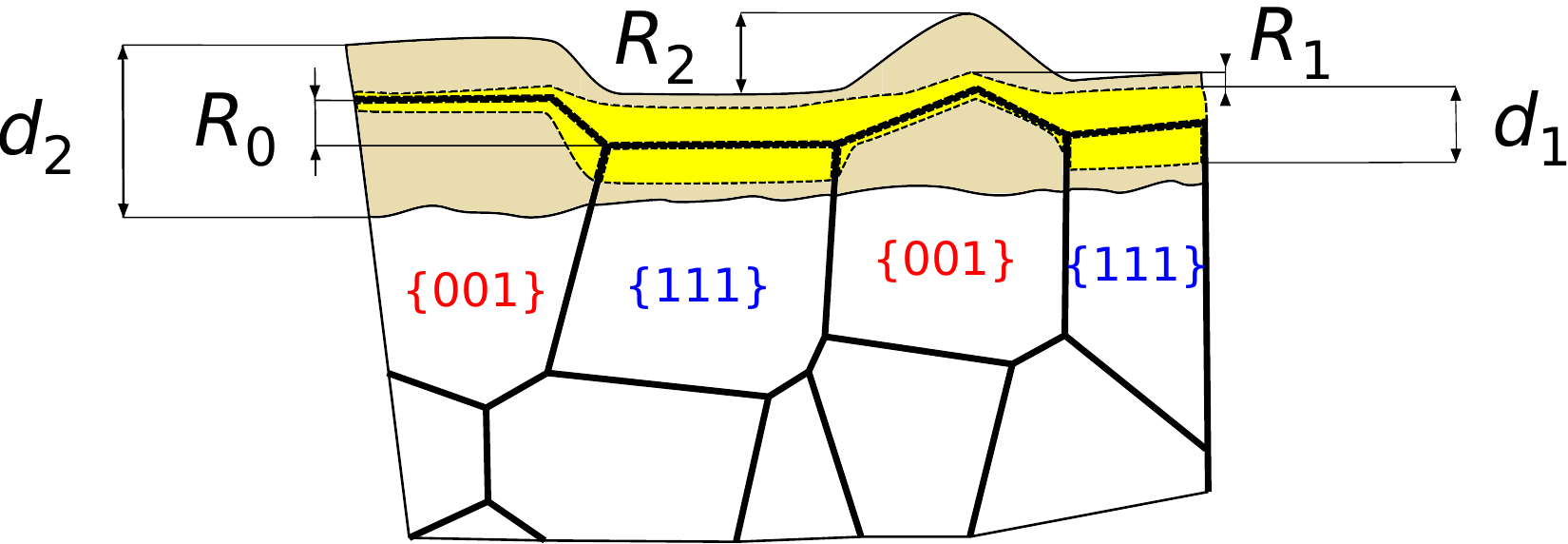}
	\caption[foo_bar]{A possible interpretation of CLSM, EBSD, and Raman spectroscopy observations showing the development of surface roughness~$R$ at different stages of oxidation. As-fabricated tungsten samples~(surface roughness~$R_0$) have softer~\hkl{111} grains, after polishing developing areas of depression on the surface, while the harder~\hkl{001} grains manifest in protrusions. At the initial stage of oxidation~($t \leq 20$~min), a higher oxide thickness~$d_1$ is observed on~\hkl{111} substrate orientations, which leads to a surface roughness~$R_1 < R_0$. At late oxidation stages~($t>72$~h), a thicker oxide~($d_2$) is observed on~\hkl{001} grains and an increased surface roughness~$R_2 > R_1$ is measured.}
	\label{fig:oxidation_drawing}
\end{figure}

A scheme~(Fig.~\ref{fig:oxidation_drawing}) illustrates the effect of oxidation on surface roughness.
The secondary electron images of as-polished samples~(Fig.~\ref{fig:sem_example}) provide information on the topography and, to a lesser degree than back-scatter signal, compositional~$Z$--contrast~\cite{Seiler1983}. In this study, as-polished samples were of $99.97$~wt.~\% purity and therefore little to no compositional contrast was observed, meaning that dark~(lower secondary electron yield) and light regions~(higher secondary electron yield) are topologically different. Darker areas with weaker signal correspond to deeper surface layers~(depressions) while bright areas indicate protrusions. Due to mechanical anisotropy~\cite{GrainPolishing1}, depressions and protrusions are observed respectively on~\hkl{111} and~\hkl{001} base material orientations~(illustrated in fig.~\ref{fig:oxidation_drawing}).

The variation in topography as characterised by the interquartile range~(IQR), which is robust to outliers, was observed to decrease at shorter oxidation runs~(Fig.~\ref{fig:Topography_histogram_plot}). This is possible only if the surface regions of depression, which correspond to~\hkl{111} type orientations, accumulate a thicker oxide -- as proposed in Sect.~\ref{sect:maps}. For longer oxidation runs, an increase in the IQR roughness~(Fig.~\ref{fig:Topography_histogram_plot}) was observed, indicating a change in the oxide layer growth kinetics with the oxide grown on~\hkl{001} substrate orientations becoming dominant over the oxide on~\hkl{111} regions -- as expected~\cite{Preferential1}. 

\section{Conclusion}

Short~(up to $72$~h) exposure of tungsten~(candidate plasma-facing material) to air or steam at~$T=400$~{\textdegree}C is a possible scenario in a fusion reactor. Under these conditions, oxidation kinetics correlate with the tungsten substrate orientation, with different oxide phases being formed in the process. Evidence was found in support of a thicker oxide on~\hkl{111} substrate orientations. On the other hand, published research reports oxidation on~\hkl{111} substrate orientations to be the slowest. At first, this seems to contradict the previous statement. However, at longer oxidation runs a change in the phase composition of the oxide layers has been shown to affect the oxide growth kinetics. This could later result in preferential oxidation on substrate orientations closer to~\hkl{001}. Hence, it can be inferred that the crystallographic preference of oxidation changes dynamically.

A robust method for smoothing, baseline subtraction and deconvolution based on a statistical information criterion was adopted to analyse the Raman scattering response of these oxides. Using this method, the formation of $h$--$\textrm{WO}_{3}$ phase was predicted in the proximity of~\hkl{001} substrate orientations just at the start of oxidation. This phase is later either substituted or covered by an $o$--WO$_3$ oxide layer. A stable orthorhombic oxide phase~($o-$WO$_3$) is reported for~\hkl{111} type substrate orientations for chamber pressures up to $0.2$~bar. At atmospheric pressure, a mixture of~$m-$WO$_2$ and~$o-$WO$_3$ is suspected.

More experiments are needed to explain the growth kinetics of tungsten oxides for thicker layers, where compositional variations are evidenced by a change in oxide colours froxm yellow to blue. Under these conditions, additional phase transformations could take place. Further work will involve compositional and structural analyses at late oxidation stages. In addition, the temperature effect on the oxide phase evolution will be studied.

\section*{Acknowledgements} 

\noindent This work was funded by the RCUK Energy Programme~(Grant No. EP/P012450/1). The research used UKAEA’s Materials Research Facility, which has been funded by and is part of the UK’s National Nuclear User Facility and Henry Royce Institute for Advanced Materials. The authors would like to thank Mr. Christopher Smith for his help in configuring the first Raman measurements and for valuable suggestions used to drive this work further.

\section*{Data availability} 

\noindent The raw/processed data required to reproduce these findings is stored in a dropbox repository and may be shared on request by contacting the corresponding author. Due to technical limitations, it is not possible to upload all data to a public repository. A software toolkit implementing the methodology described for data processing in this work may also be made available on request.

\bibliography{Big_Bibliography}

\end{document}